\pdfoutput=1
\documentclass[manuscript,screen]{acmart}

\AtBeginDocument{%
  \providecommand\BibTeX{{%
    Bib\TeX}}}

\setcopyright{acmcopyright}
\copyrightyear{2023}
\acmYear{2023}
\acmDOI{XXXXXXX.XXXXXXX}
\acmBooktitle{ACM Transactions on Architecture and Code Optimization}

\usepackage{graphicx}
\usepackage{xspace}
\usepackage{xcolor}
\usepackage{textcomp}
\usepackage{listings}
\usepackage{verbatim}

\usepackage[export]{adjustbox}
\usepackage{listings}
\usepackage[frozencache=true,cachedir=minted-cache]{minted}
\definecolor{LightGray}{gray}{0.9}

\def\BibTeX{{\rm B\kern-.05em{\sc i\kern-.025em b}\kern-.08em
    T\kern-.1667em\lower.7ex\hbox{E}\kern-.125emX}}

\newcommand{\gemm}{\textsf{GEMM}\xspace}
\newcommand{\PF}{\textsf{PFACT}\xspace}
\newcommand{\solve}{\textsf{TSOLVE}\xspace}

\definecolor{gray98}{rgb}{0.93,0.93,0.93}
\definecolor{gray20}{rgb}{0.20,0.20,0.20}
\definecolor{gray25}{rgb}{0.25,0.25,0.25}
\definecolor{gray16}{rgb}{0.161,0.161,0.161}
\definecolor{gray60}{rgb}{0.6,0.6,0.6}
\definecolor{gray30}{rgb}{0.3,0.3,0.3}
\definecolor{bgray}{RGB}{248, 248, 248}
\definecolor{amgreen}{RGB}{40, 144, 40}
\definecolor{myblue}{RGB}{0, 40, 255}
\definecolor{amred}{RGB}{228,26,28}
\definecolor{amethyst}{rgb}{0.6, 0.4, 0.8}
\definecolor{mymauve}{rgb}{0.58,0,0.82}
\definecolor{LightGray}{gray}{0.9}
\definecolor{mygray}{rgb}{0.95,0.95,0.95}

\begin{document}
\title{Co-Design of the Dense Linear Algebra
       Software Stack for 
       Multicore Processors}

\author{Héctor Martínez}
\orcid{0000-0001-5891-4479}
\email{el2maph@uco.es}
\affiliation{%
  \institution{Universidad de Córdoba}
  \city{Córdoba}
  \country{Spain}
}
\author{Sandra Catalán}
\orcid{0000-0002-9321-2728}
\email{catalans@uji.es}
\affiliation{%
  \institution{Universitat Jaume~I}
  \city{Castellón de la Plana}
  \country{Spain}
}
\author{Francisco D. Igual}
\orcid{0000-0003-4480-9517}
\email{figual@ucm.es}
\affiliation{%
  \institution{Universidad Complutense de Madrid}
  \city{Madrid}
  \country{Spain}
}
\author{José R. Herrero}
\orcid{0000-0002-4060-367X}
\email{josep@ac.upc.edu}
\affiliation{%
  \institution{Universitat Politècnica de Catalunya}
  \city{Barcelona}
  \country{Spain}
}
\author{Rafael Rodríguez-Sánchez}
\orcid{0000-0001-8789-3953}
\email{rafael.rodriguez@uclm.es}
\affiliation{%
  \institution{Universidad de Castilla-La Mancha}
  \city{Albacete}
  \country{Spain}
}
\author{Enrique S. Quintana-Ortí}
\orcid{0000-0002-5454-165X}
\email{quintana@disca.upv.es}
\affiliation{%
  \institution{Universitat Politècnica de València}
  \city{Valencia}
  \country{Spain}
}

\renewcommand{\shortauthors}{Martínez et al.}

\begin{abstract}
This paper advocates for an intertwined design of the dense linear algebra software stack that breaks down the strict barriers between the high-level, blocked algorithms in LAPACK (Linear Algebra PACKage) and the low-level, architecture-dependent kernels in BLAS (Basic Linear Algebra Subprograms). Specifically, we propose customizing the \gemm (general matrix multiplication) kernel, which is invoked from the blocked algorithms for relevant matrix factorizations in LAPACK, to improve performance on modern multicore processors with hierarchical cache memories. To achieve this, we leverage an analytical model to dynamically adapt the cache configuration parameters of the \gemm to the shape of the matrix operands. Additionally, we accommodate a flexible development of architecture-specific micro-kernels that allow us to further improve the utilization of the cache hierarchy.

Our experiments on two platforms, equipped with ARM (NVIDIA Carmel, Neon) and x86 (AMD EPYC, AVX2) multi-core processors, demonstrate the benefits of this approach in terms of better cache utilization and, in general, higher performance. However, they also reveal the delicate balance between optimizing for multi-threaded parallelism versus cache usage.
\end{abstract}



\maketitle              

\section{Introduction}

\begin{figure}[thb!]
\centering
\includegraphics[width=0.9\textwidth]{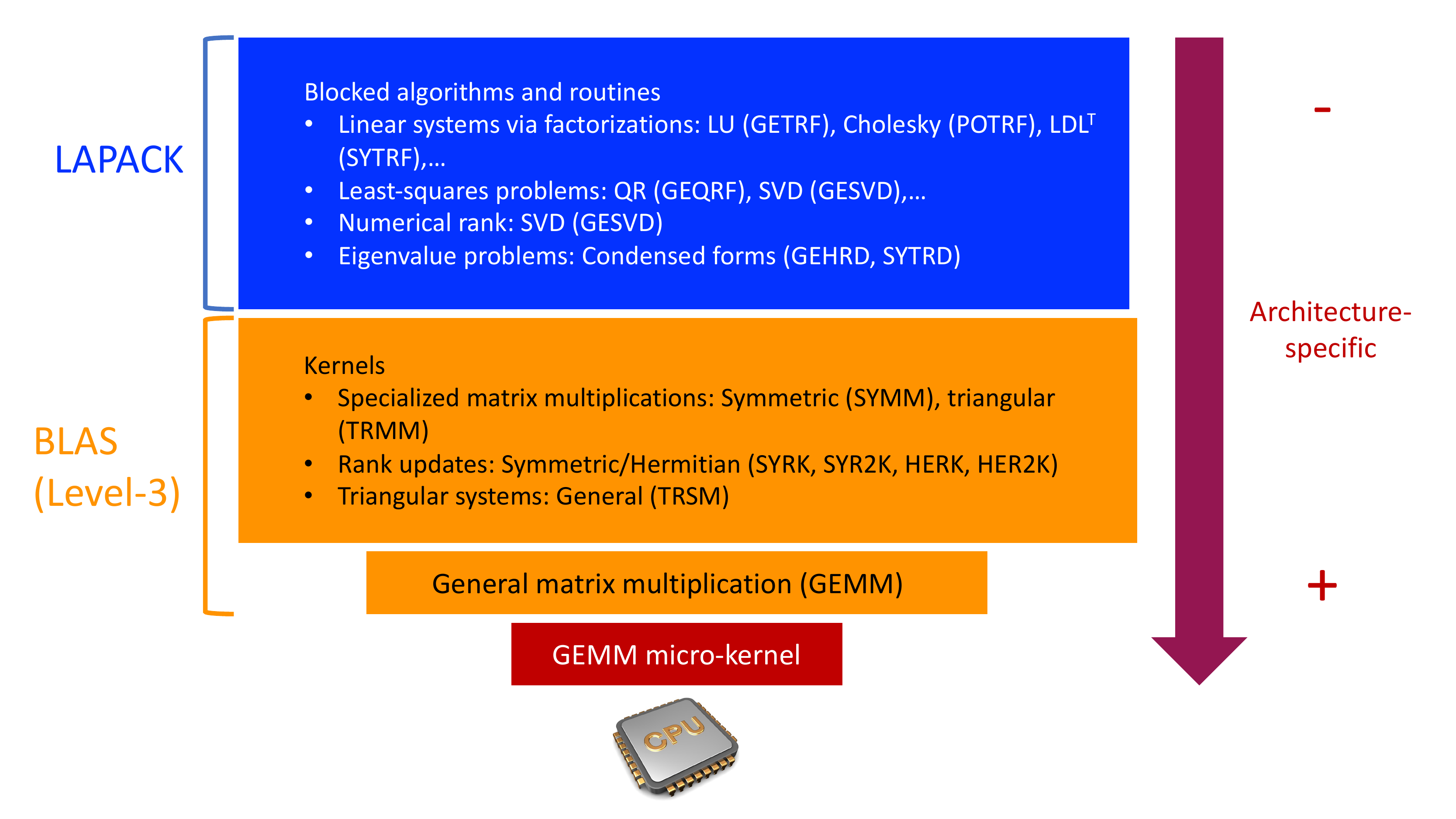}
\vspace*{-2ex}
\caption{Dense linear algebra software stack on top of processor architecture.}
\label{fig:stack}
\end{figure}

LAPACK (Linear Algebra PACKage)~\cite{lapack} is a library 
that comprises a large collection of numerically-reliable routines, with a well-defined interface,
for tackling the linear systems, least-squares 
problems, numerical rank
computations, and eigenvalue problems that arise in many scientific 
applications~\cite{doi:10.1137/1.9781611971446,GVL3,GO91}. 
To achieve high performance, whenever possible, the
LAPACK routines are formulated as blocked algorithms~\cite{DonDSV98}, 
where a significant portion of their (arithmetic) computations 
is performed via a reduced number of computational kernels
known as the Level-3 BLAS (Basic Linear Algebra Subprograms)~\cite{BLAS3}.
In addition, for portability, a majority of the Level-3 BLAS are built on top of the
general matrix multiplication kernel (\gemm)~\cite{10.1145/292395.292412,Goto:2008:HIL3}, which
is also part of the Level-3 BLAS. 
Finally, most high performance realizations of \gemm cast their arithmetic in
terms of a small, processor-specific component,
known as the micro-kernel,
which is usually written in assembly~\cite{Goto:2008:AHP,BLIS1,OpenBLAS}.
The organization of the dense linear algebra (DLA) software stack is illustrated in
Figure~\ref{fig:stack}, displaying four levels (boxes): LAPACK 
(blocked) algorithms, Level-3 BLAS kernels, Level-3 BLAS \gemm,
and micro-kernel.
From top to bottom, the algorithms/routines/kernels become more architecture-specific.

The foundations that underlie high performance instances of \gemm, in numerical libraries such as
GotoBLAS~\cite{Goto:2008:AHP}, BLIS~\cite{BLIS1}, OpenBLAS~\cite{OpenBLAS:ICPADS}, AMD AOCL, ARMPL, and (possibly) 
Intel MKL and oneAPI,
were set  in a seminal paper from K. Goto and R. A. van de Geijn in 2008~\cite{Goto:2008:AHP}.
In general, these realizations of \gemm 
 are highly optimized for large, ``squarish'' matrix operands.
In contrast, they offer significantly lower
performance in other scenarios. 
Unfortunately, most blocked algorithms in LAPACK actually invoke
the \gemm kernel with operands which are far from being square and, 
in many cases, where at least one of the three dimensions of the matrix multiplication is small.

In more detail, in order to tackle the memory wall~\cite{10.1145/216585.216588,McKee2011}, the blocked algorithms for
most matrix factorizations in LAPACK iteratively
``process'' the data in panels of $b$ columns per iteration. 
This parameter $b$ is known as the algorithmic block size  and,
as will be discussed later, it
is linked with one of the dimensions of the matrix multiplication.
In practice, inside the algorithm for the matrix
factorization $b$ is set to a small value,
in order to maximize the amount of floating point arithmetic operations (flops) cast in terms of \gemm.
However, this produces the non-square \gemm kernels
with the potential low performance. 
While some instances of BLAS are furnished with 
specialized routines to deal with 
this type of matrix operands, they are still built on top of a single micro-kernel, which we will show
is also a source of low performance.

In dealing with this problem, the paper makes the following
specific contributions:

\begin{enumerate}
\item We identify the cause of the low performance of \gemm for the cases
      arising in blocked algorithms for matrix factorizations in the inefficient
      utilization of the cache hierarchy, in particular the L2 cache.
\item We provide a practical demonstration that
      the problem can be overcome by applying two techniques:
\begin{enumerate}
\item Carefully adjust the cache configuration parameters for \gemm, 
which determine the cache usage  efficiency, by leveraging 
a refined analytical model that guides their selection,
as well as providing a sufficiently-large workspace 
buffers to \gemm.
\item Develop tailored realizations of the 
micro-kernel using either vector intrinsics 
assisted with a modern compiler~\cite{Ala22} or directly in assembly. 
\end{enumerate}
\item For a multi-threaded execution, we expose the delicate balance between 
      improving the cache usage and accommodating 
      a higher degree of parallelism.
\item We demonstrate the benefits of this approach
      using the sequential and 
      multi-threaded execution of the LU factorization
      on ARM- and AMD-based multicore platforms.
\end{enumerate}

The previous contributions and experimental analysis support the idea that lowest two levels in the 
DLA stack (specifically, the matrix-matrix multiplication upon which the rest of the stack is built), 
should be designed and built in close coordination 
with the upper levels of the stack, following a {\em co-design approach}. 
Specifically, our work conveys the following message which deviates from the current approach:
For high performance, the implementation of \gemm 
should not statically select the cache configuration parameters or the most appropriate micro-kernel. 
Instead, the dynamic selection of these two ``components'' should be performed at runtime using analytical models that consider both the cache hierarchy and operand shapes. Ultimately, the selection should depend on the specific operation and operand shapes dictated by the algorithm in the upper levels of the DLA stack.

The rest of the paper is structured as follows.
In Section~\ref{sec:stack}
we briefly review the DLA software stack using the
LU factorization as a use case,
and connecting this algorithm to \gemm. 
Then, in Section~\ref{sec:optimization},
we describe the modern realization of \gemm,
targeting an (ARM-based) NVIDIA Carmel core for illustrative
purposes.
In Section~\ref{sec:experiments}
we provide strong experimental evidence of the benefits
of our approach, and the sources of the performance
gains on two platforms, respectively with ARM and AMD
technologies, 
leveraging hardware counters when available. 
Finally, in Section~\ref{sec:remarks} we close the paper
with a few concluding remarks.

\section{Review of the DLA Software Stack via the LU Factorization}
\label{sec:stack}

In this section, we illustrate the interplay between the LAPACK-level blocked algorithms,
using the LU factorization~\cite{GVL3} as a case study, and the underlying \gemm kernel and micro-kernel. 
Furthermore, we review the foundations of modern realizations of matrix multiplication in current
high performance instances of BLAS.

\subsection{LU factorization}

\newcommand{\FlaThreeByThreeTL}[9]{
\left[
\begin{array}{c | c  c}
#1 & #2 & #3 \\ \hline
#4 & #5 & #6 \\ 
#7 & #8 & #9
\end{array}
\right]
}

Figure~\ref{fig:blocked_LU} displays a right-looking blocked algorithm that computes the
LU factorization of an $s \times s$ matrix $A$, 
expressed using a high-level Matlab-like notation with the matrix indices starting at 0. 
Assume, for simplicity, that $s$ is an integer multiple of the algorithmic block size $b$
and, in order to describe the operations performed 
at the $(k/b)$-th iteration of loop labeled as~\texttt{F1} in the algorithm, 
consider the following partitioning of the matrix:
\[
 A \rightarrow
 \FlaThreeByThreeTL
 {A_{00}}{A_{01}}{A_{02}}
 {A_{10}}{A_{11}}{A_{12}}
 {A_{20}}{A_{21}}{A_{22}}
 =
 \FlaThreeByThreeTL
 {A(l,l)}{A(l,p)}{A(l,t)}
 {A(p,l)}{A(p,p)}{A(p,t)}
 {A(t,l)}{A(t,p)}{A(t,t)}.
\]
In the expression, $A_{11} = A(p,p)$ corresponds the diagonal block consisting of the
intersection of the matrix rows and columns in the range $p=[k:k+b-1]$;
and the remaining blocks are defined by the ranges
$l=[0:k-1]$ and $t=[k+b:s-1]$.
Inside the loop body, the algorithm then initially computes the panel factorization (\PF)
\[
    \left[
    \begin{array}{c}
    A_{11}\\
    A_{21}
    \end{array}
    \right] = 
    \left[
    \begin{array}{c}
    A(p,p)\\
    A(t,p)
    \end{array}
    \right] = 
    \left[
    \begin{array}{c}
    L_{11}\\
    L_{21}
    \end{array}
    \right] U_{11}, 
\]
where $L_{11}$ is unit lower triangular and $U_{11}$ is upper triangular,
both of dimension $b \times b$,
and these triangular factors overwrite the corresponding parts of $A_{11}$.
Here the diagonal elements of $L_{11}$ are not explicitly stored as they all equal 1. 
Furthermore, $L_{21}$ overwrites $A_{21}$.  

\begin{figure}[thb]
\begin{lstlisting}[language=C,morekeywords={matrix}]
void blocked_LU( matrix A; int s, b )
{
  for (k=0; k<s; k+=b) { // Loop F1
    p  = [k:k+b-1]; // Column indices for the panel
    t  = [k+b:s-1]; // Indices for the trailing part
    // Factorize current panel
    @*$\PF$*)( [ A(p,p);                  // [A11] = [L11] * U11
             A(t,p) ] );              // [A21]   [L21]   
    // Trailing update of panels
    @*$\solve$*)( A(p,p), A(p,t) );         // U12 = L11^-1 * A12
    @*$\gemm$*)(  A(t,t), A(t,p), A(p,t) );  // A22 = A22 - L21 * U12
  }
}
\end{lstlisting}
\caption{Simplified routine for the LU factorization.}
\label{fig:blocked_LU}
\end{figure}

Following the panel factorization, the trailing update for the $(k/b)$-th iteration 
involves the triangular system solve (\solve):
\[
U_{12} = L_{11}^{-1} \cdot A_{12} \quad \equiv \quad
A(p,t) = \textsc{Lower}(A(p,p))^{-1} \cdot A(p,t), 
\]
where 
$\textsc{Lower}(\cdot)$
is an operand that assembles a unit lower triangular factor ($L_{11}$) from the strictly lower
part of the input operand;
and the result $U_{12}$ overwrites the contents of $A_{12}$.
This is then
followed by the \gemm:
\[
A_{22} = A_{22} - L_{21} \cdot U_{12} \quad \equiv \quad A(t,t) = A(t,t) - A(t,p) \cdot A(p,t).
\] 
For brevity, we omit partial pivoting~\cite{GVL3}
from the discussion of the algorithm, though
all our implementations include this technique.

Note the variation in the dimensions of the operands of
\PF, \solve, \gemm as the factorization progresses. With respect to this,
the matrix multiplication updates a square block, corresponding
to the trailing part of the matrix, $A_{22}$, 
with the product of two narrow panels: $A_{21}$ and $A_{12}$, 
with column and row dimension $b$, respectively.
From the performance viewpoint, 
\PF is mostly sequential and stands in the critical path of the algorithm,
while 
\solve and \gemm exhibit a fair amount of thread-level parallelism.
This exposes the role of the algorithmic block size $b$ on the performance of the blocked
algorithm for the LU factorization:
On the one hand, increasing $b$ 
shifts the computational workload 
from the efficient, parallel kernels in the trailing update toward the mostly-sequential \PF.
On the other hand,  choosing a small $b$ reduces
the ratio between arithmetic and memory accesses for \gemm in the trailing update,  
lowering the arithmetic throughput of this kernel.
To balance both factors, in 
practice the algorithmic block size is chosen in the range $b\in [64,384]$. 

\subsection{High performance, multi-threaded GEMM}

\begin{figure}[thb!]
\centering
\begin{tabular}{ccc}
\begin{minipage}[c]{0.46\textwidth}
\footnotesize
\begin{lstlisting}[language=C,morekeywords={matrix}]
void blocked_GEMM( matrix C, A, B; 
                   int m, n, k, 
                       mc, nc, kc, mr, nr )
{
  for (jc=0; jc<n; jc+=nc)           // Loop G1
    for (pc=0; pc<k; pc+=kc){        // Loop G2
      // Pack B
      Bc:=B(pc:pc+kc-1, jc:jc+nc-1);
      for (ic=0; ic<m; ic+=mc){      // Loop G3
        // Pack A
        Ac:=A(ic:ic+mc-1, pc:pc+kc-1);
        for (jr=0; jr<nc; jr+=nr)    // Loop G4
          for (ir=0; ir<mc; ir+=mr)  // Loop G5
            mukernel( C(ic+ir:ic+ir+mr-1, 
                        jc+jr:jc+jr+nr-1), 
                      Ac(ir:ir+mr-1,0:kc-1), 
                      Bc(0:kc-1,jr:jr+nr-1), 
                      mr, nr, kc );
      } 
    } 
}
\end{lstlisting}
\vspace*{12ex}
\end{minipage}
& ~~~ &
\begin{minipage}[c]{0.44\textwidth}
\begin{tabular}{l}
\footnotesize
\begin{lstlisting}[language=C,alsoletter={.},deletekeywords={.sum},morekeywords={matrix}]
void mukernel( matrix Cr, Ar, Br; 
               int mr, nr, kc )
{
  for (pr=0; pr<kc; pr++) // Loop M1
    Cr += Ar(0:mr-1,pr) 
       *  Br(pr,0:nr-1);
}
\end{lstlisting}
~\\
~\\
\includegraphics[width=\textwidth]{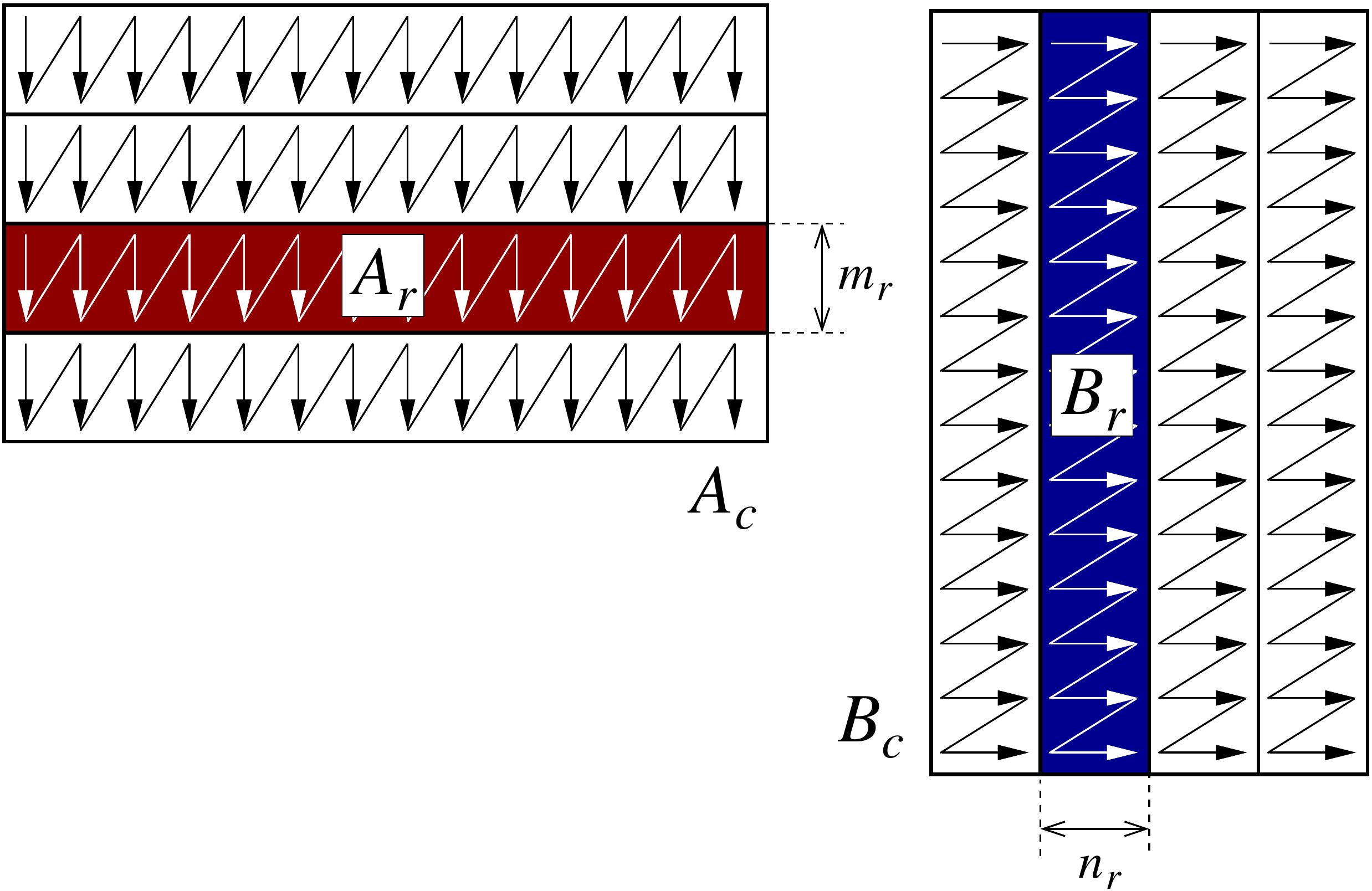}
\end{tabular}
\end{minipage}
\end{tabular}
\caption{High performance algorithm for \gemm. Left: Blocked algorithm; Top-Right: Micro-kernel; Bottom-Right: Packing.}
\label{fig:blocked_GEMM}
\end{figure}

Consider now the \gemm
$C = A\cdot B$,  where 
$A \rightarrow m \times k$,
$B \rightarrow k \times n$, and 
$C \rightarrow m \times n$.
The implementation of this kernel in modern libraries (e.g.,
BLIS, OpenBLAS, AMD AOCL, ARM PL and, 
possibly, Intel oneMKL) follows
GotoBLAS2~\cite{Goto:2008:AHP} to
formulate it as a blocked algorithm, consisting of
five nested loops 
around two packing routines and a 
\textit{micro-kernel}; see Figure~\ref{fig:blocked_GEMM}, left,
and the loops labeled as \texttt{G1}, \texttt{G2},\ldots, \texttt{G5} there.
(For simplicity, hereafter we assume that 
$m,n,k$ are integer multiples of $m_c,n_c,k_c$, respectively; and 
$m_c,n_c$ are integer multiples of $m_r,n_r$, respectively).

\begin{figure}[t]
\centering 
\begin{tabular}{c}
\includegraphics[width=0.5\textwidth]{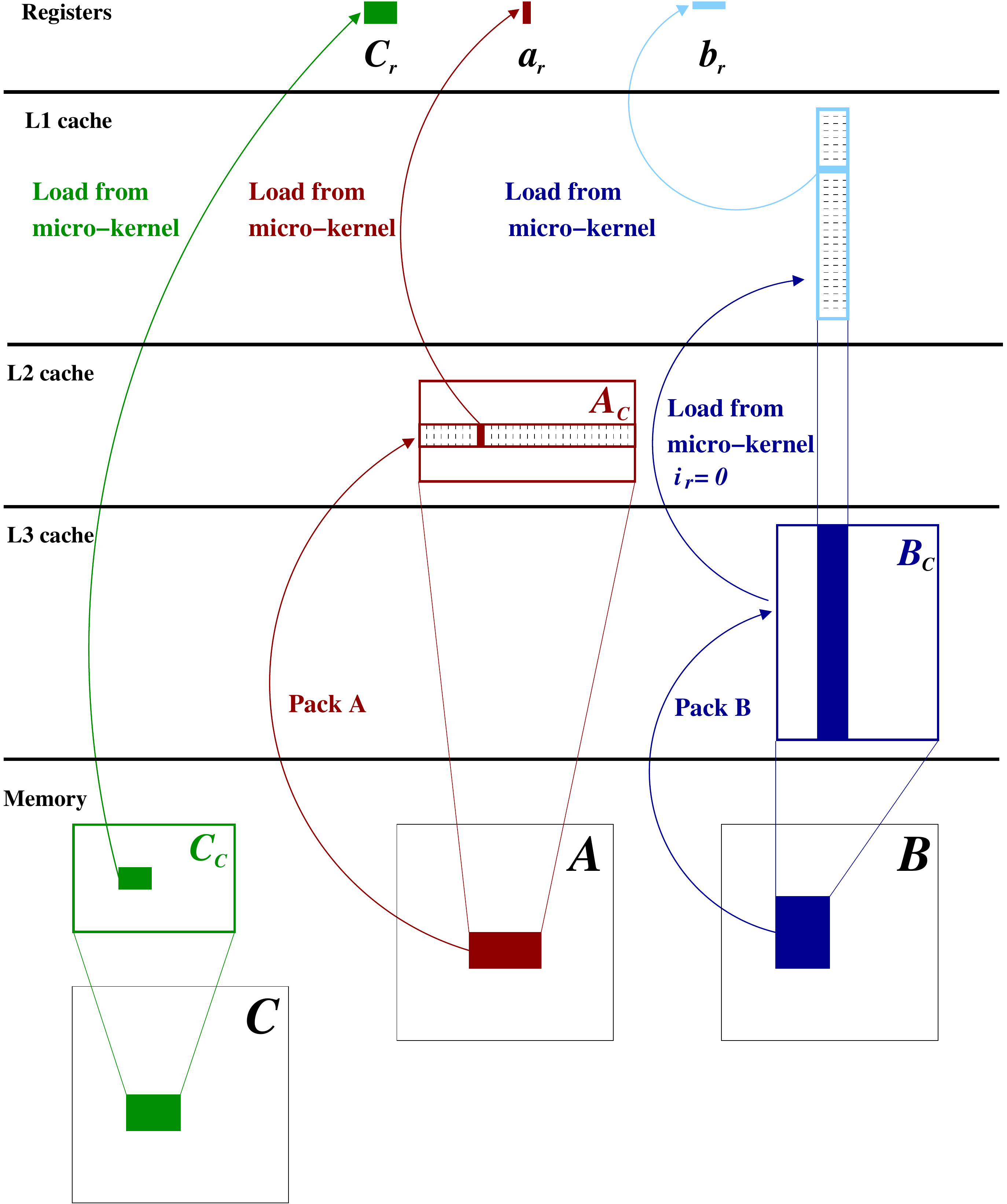}
\end{tabular}
\caption{Data movement in the modern implementations of \gemm.}
\label{fig:blis_movement}
\end{figure}

For the blocked realization of the \gemm kernel in the figure, 
an appropriate choice of the strides 
for the three outermost loops,
given by $m_c,n_c,k_c$ and known as the \textit{cache configuration parameters} (CCPs),
combined with a certain packing of the matrix inputs into
two buffers, $A_c \rightarrow m_c \times k_c$
and $B_c \rightarrow k_c \times n_c$,
(see Figure~\ref{fig:blocked_GEMM}, bottom-right)
orchestrates a careful pattern of data movements across the memory hierarchy that
reduces the cache miss rate;
see Figure~\ref{fig:blis_movement}.
In particular, the blocked algorithm aims at maintaining
a $k_c \times n_r$ micro-panel of $B_c$ 
(like that labeled as $B_r$ and highlighted in blue in Figure~\ref{fig:blocked_GEMM}, bottom right)
in the L1 cache during the execution
of loop \texttt{G5}; 
the $m_c \times k_c$ buffer $A_c$ in the L2 cache during 
loop \texttt{G4}; 
and the $k_c \times n_c$ buffer $B_c$ in the L3 cache during 
loop \texttt{G3}~\cite{BLIS1,BLIS4}. 

Due to the lack of dependencies, parallelizing the \gemm kernel for a multicore processor 
is basically attained by
exploiting loop-parallelism at one or more of the loops 
\textsf{G1},
\textsf{G3} and/or
\textsf{G4}.
Parallelizing loop \textsf{G2} should be avoided because of write-after-write (WAW) dependencies~\cite{HenP17} 
that yield
a race condition, while 
loop \textsf{G5} is usually too fine-grain to yield an efficient parallelization on a multicore processor.
 The multi-threaded parallelization of the \gemm kernel has been previously analyzed for conventional multicore processors~\cite{BLIS2}, modern many-threaded architectures~\cite{BLIS3}, and low-power (asymmetric) ARM-based processors~\cite{Catalan2016}.
 The general conclusion from these analyses is that, when the L1 cache is private and
 the L2 cache is shared, in general the best option is to parallelize loop G4. 
 When both the L1 and L2 cache levels are private, then loop G3 is usually a better choice.
 Finally, when the number of cores is large, it may be more beneficial to parallelize
 two (or more) loops.

\subsection{Architecture-specific micro-kernel}

The \gemm
micro-kernel is usually written in assembly and, therefore, it is an architecture-specific piece of code. It comprises a single loop
(labeled as \texttt{M1} in Figure~\ref{fig:blocked_GEMM}, top-right) 
that performs a sequence of $k_c$ rank-1 updates on an $m_r \times n_r$ 
\textit{micro-tile} of
$C$, denoted as $C_r$, each involving 
a column of an $m_r \times k_c$ micro-panel of 
the packed buffer
$A_c$ and a row of a $k_c \times n_r$ micro-panel of the packed buffer $B_c$.
(See the blocks labeled as $A_r$ and $B_r$ in Figure~\ref{fig:blocked_GEMM}, bottom-right.)

\newcommand{\mk}{\textsf{MK}}
\newcommand{\mkflp}{\mk_{\textrm{flp}}}
\newcommand{\mkCmem}{\mk_{\textrm{Cmem}}}
\newcommand{\mkAmem}{\mk_{\textrm{Amem}}}
\newcommand{\mkBmem}{\mk_{\textrm{Bmem}}}

The micro-kernel dimensions, $m_r,n_r$, should be chosen
as large as possible without incurring into register spilling~\cite{Dowd98}.
Furthermore,  it is convenient that $m_r\approx n_r$
as this maximizes the ratio of {\em flops} 
(floating point operations) to 
{\em memops} (memory accesses) 
performed by the micro-kernel, given by
\[
\begin{array}{l}
\mkflp /(\mkCmem + \mkAmem + \mkBmem) =\\[0.1in]
  \hspace*{16ex} 2m_rn_rk_c / (2m_rn_r+m_rk_c+k_cn_r).
\end{array}
\]
Note that the micro-tile $C_r$ is read from memory into the processor registers
before the micro-kernel commences its execution and written back into the memory once
it is complete. This accounts for the factor 2 in the $\mkCmem$ term as we make
no distinction between the costs of data reads and writes.
In line with this simplification, we observe that, during the micro-kernel execution,
the memory accesses to the different matrix operands can be expected to
encounter the data in distinct levels of the memory hierarchy and, therefore, 
not all of the accesses will present the same cost.

For processors equipped with SIMD (single instruction, multiple data) 
units, one of the micro-kernel dimensions $m_r$ or $n_r$,
is chosen to perform the accumulation on $C_r$ 
using SIMD arithmetic instructions; 
and the special packing of the elements of $A,B$ into the buffers $A_c,B_c$ 
enables loading their data using SIMD instructions.

\vspace*{1ex}
\noindent
\textbf{Summary:}
The previous review of the 
DLA software stack, using the LU factorization as a 
case study, can be condensed into the following observations:
\begin{itemize}
\item The performance of the
      blocked algorithm for the matrix factorization is strongly dictated by the
      parameter $b$, which controls the balance between
      the flops in the mostly-sequential panel factorization versus those in the parallel trailing update.
      Furthermore, $b$ has a direct impact on the performance of the \gemm in the trailing update, which benefits from
      a large value for that parameter.
    
\item The blocked algorithm for \gemm is parameterized by the tuple 
       ($m_c,n_c,k_c$), which has to be adjusted to fit the cache
       system of the target architecture~\cite{BLIS4}.
       For the type of \gemm kernels arising in the LU factorization (and other matrix decompositions), $k_c \le b$.
\item For high performance, the micro-kernel is usually written in assembly and, therefore,
       it is architecture-specific. The dimensions of the micro-kernel, given by $m_r \times n_r$,
       are usually set to maximize the flops/memops ratio 
       without incurring into register spilling.
\end{itemize}
\section{Optimization of the Cache Utilization for \gemm}
\label{sec:optimization}

\newcommand{\mkdim}[2]{\mk_{#1\times#2}}

The review of the blocked algorithm for the LU factorization in the previous section exposes 
the link between the algorithmic block size $b$ and the
dimension $k$ of the \gemm that dominates the arithmetic cost of the trailing
update. 
In this section, we discuss the effect of this parameter on the cache usage and
hint on the subsequent impact on performance.

For all the discussions and experiments in the paper, 
we consider IEEE 64-bit floating point arithmetic (FP64).
Also, the performance results correspond to average numbers 
collected for a large number of repetitions of
each experiment.
When necessary, we will specify the dimensions of the micro-kernel, $m_r \times n_r$,
using the notation 
$\mkdim{m_r}{n_r}$.

\subsection{Target architecture and \gemm realization}

\begin{figure}[t]
\centering 
\begin{tabular}{c}
\includegraphics[width=0.8\textwidth]{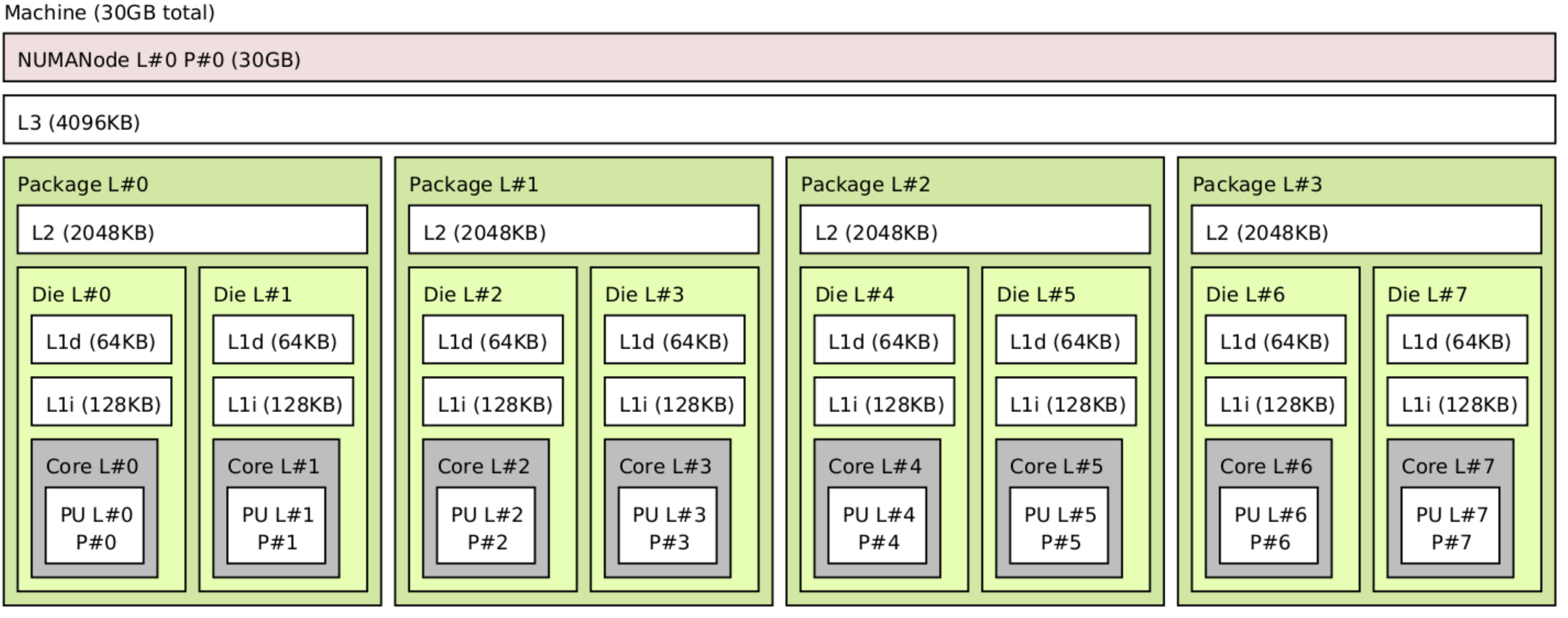}
\end{tabular}
\caption{Memory hierarchy of the NVIDIA Carmel processor determined using \texttt{hwloc}.}
\label{fig:nvidia_carmel}
\end{figure}

For illustrative purposes and brevity, 
in the remainder of this section
we target a single core of the NVIDIA Carmel processor (ARMv8.2)
on an NVIDIA Jetson AGX Xavier board.
Each core in this platform operates a 4-associative L1 cache 
with a capacity of 64~KB; shares 
a 16-associative L2 cache of 2~MB with a sibling core;
and shares a 16-way associative L3 cache of 4~MB with the 
remaining 7 cores of the processor; see Figure~\ref{fig:nvidia_carmel}.
The cores are equipped with 128-bit SIMD arithmetic units.

In addition, 
we consider the high performance realization of \gemm in BLIS (version 0.8.1).
For the ARMv8.2 architecture and FP64, 
BLIS 
encodes a micro-kernel with $m_r\times n_r = 6 \times 8$ and sets
the CCPs as
\[
(m_c^B,n_c^B,k_c^B) = (120,3072,240).
\]
Note that, \textit{even though $k_c^B=240$ for BLIS, 
the actual CCP is given by $k_c=\min(k,k_c^B)$}.
(A similar remark applies to the actual $m_c,n_c$.)

\subsection{Utilization of the cache hierarchy}

The blocked algorithm for \gemm in BLIS (and all other libraries which follow the 
GotoBLAS2 scheme)
is designed in order to favor that, during the execution of the micro-kernel, the data for the operands $B$ and $A$ 
are fetched from
a 
$k_c \times n_r$ 
micro-panel $B_r$
that resides in the L1 cache 
and an
$m_c \times k_c$ 
buffer $A_c$ 
in the L2 cache~\cite{BLIS1}. 
\textit{All the cache occupancy rates given next,
for $B_r$ in the L1 cache and $A_c$ in the L2 cache, 
are theoretical values 
derived from the dimensions of these two data blocks and the CCPs.}

\begin{figure}[t]
\normalsize
\centering
\begin{tabular}{ccc}
\begin{minipage}[c]{0.30\textwidth}
{
\setlength{\tabcolsep}{4pt}
\begin{tabular}{|r||rr|rr|}
\hline
$k$  & L1   &      & L2 & 
     \\
     & KB   & (\%) & KB &  (\%) 
     \\ \hline \hline

      64&        4.0&     6.2 &           60.0&     2.9 \\ 
      96&        6.0&     9.4 &           90.0&     4.4 \\ 
     128&        8.0&    12.5 &          120.0&     5.9 \\ 
     160&       10.0&    15.6 &          150.0&     7.3 \\ 
     192&       12.0&    18.7 &          180.0&     8.8 \\ 
     224&       14.0&    21.9 &          210.0&    10.3 \\ 
     240&       15.0&    23.4 &          225.0&    11.0 \\ 
     2000&      15.0&    23.4 &          225.0&    11.0 \\ 
     \hline
\end{tabular}
}
\end{minipage}
& ~~~ &
\begin{minipage}[c]{0.54\textwidth}
\includegraphics[width=1.0\textwidth]{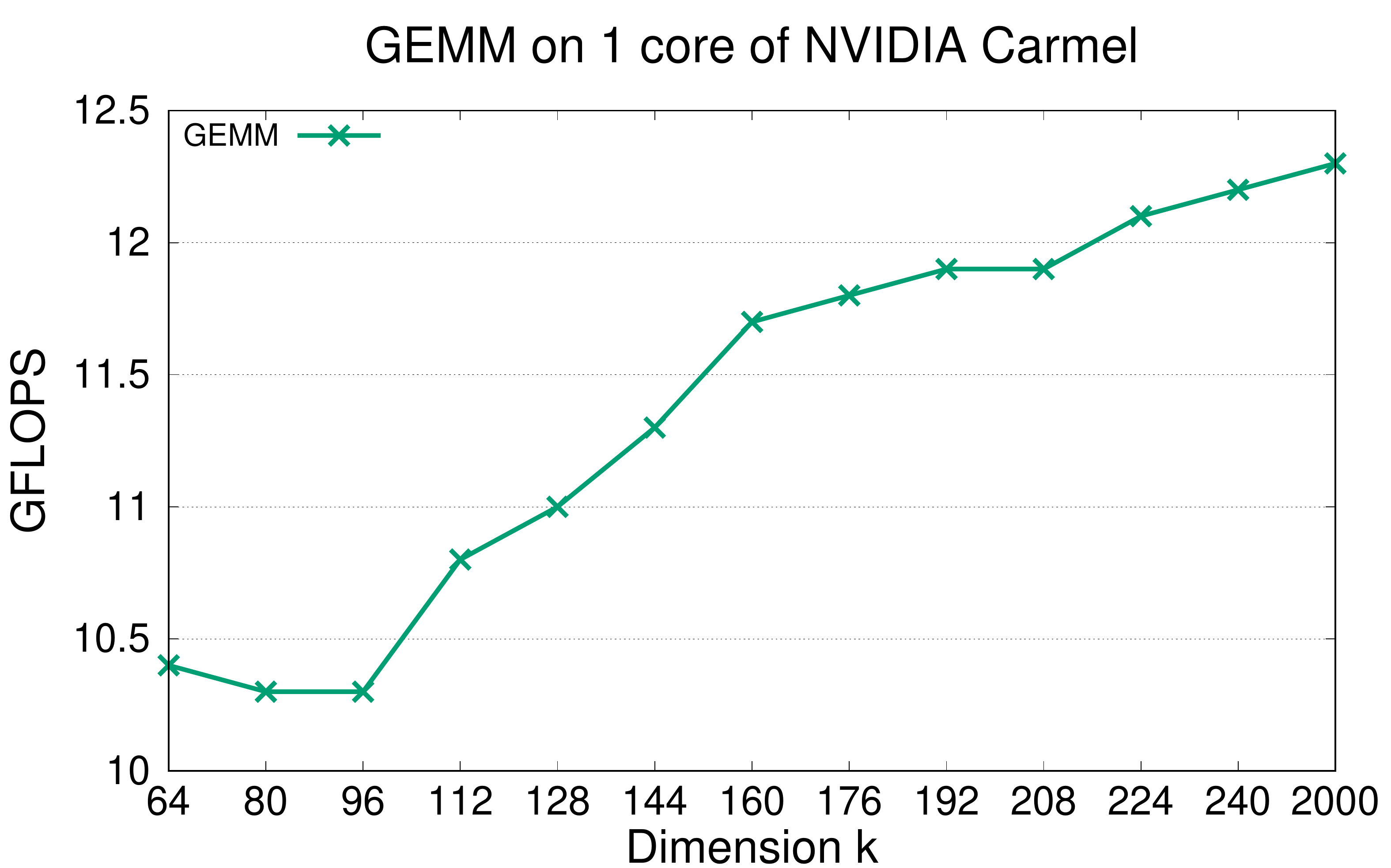}
\end{minipage}
\end{tabular}
\caption{Evaluation of the \gemm in BLIS for problems of dimension
         $m=n=2000$ and varying $k$ on a single core of an NVIDIA Carmel processor. Left:
         Theoretical cache occupation of $B_r$\textbar$A_c$ in the L1\textbar L2 caches, taking
         into account the dimensions of these blocks and those of the \gemm.
         Right: Experimental performance in GFLOPS (billions of flops per second).}
   
\label{fig:cache_BLIS}
\end{figure}

Let us first study 
the interplay between $k_c$, $k$, 
and the (theoretical) utilization
of the cache memory by
the blocked algorithm for \gemm. For simplicity, we focus on the top two levels
of the cache hierarchy only. 
For the BLIS CCPs,
Figure~\ref{fig:cache_BLIS} (left) shows the relation between 
the $k$-dimension of $B_r$\textbar $A_c$,
and the occupancy of the L1\textbar L2 caches with these data
for \gemm kernels of dimension $m=n=2000$
with $k \in \{[64,240],2000\}$.
The table shows that increasing $k$ up to
$k_c^B=240$,
results in a 
higher utilization of the L1 cache by $B_r$ (up to 23.4\%) and the L2 cache by $A_c$ (up to 11.0\%). 
From that point on, a larger $k$ does not increase the
occupation of the L1 and L2 caches since $k_c=\min(k_c,k_c^B)$.

To put this ``low'' utilization of the cache into perspective, the analytical model 
in~\cite{BLIS4} specifies that, for this particular architecture and dimensions of the micro-kernel,
$B_r$ should at most occupy 33.2\% of the L1 cache, while
$A_c$ could utilize up to 87.5\% of the L2 cache.
The reason is that,
for a 4-way associative L1 cache and the BLIS micro-kernel $\mkdim{6}{8}$,
the model indicates that
one line of each cache set should be dedicated to $C$, while the remaining lines should
be distributed between the entries of $B$ and $A$ proportionally to the ratio $n_r/m_r = 8/6 \approx 1.33$. Therefore, we can consign 
up to 32~KB 
(50\%) of the L1 cache
to $B_r$.
Similarly, for the 16-way L2 cache, 
one line of each cache set is reserved for $C$, and the remaining lines
are to be distributed between $A$ and $B$ following a ratio $k_c/n_r=240/8=30$. Thus,
we consign 14 lines per set to $A$, 
yielding a maximum usage  of 1.75~MB (87.5\%) of the L2 cache for $A_c$.

The previous discussion exposes that the CCPs selected by BLIS are far from yielding
an optimal utilization of the cache hierarchy.
The practical consequence of this issue is illustrated in
the right plot in Figure~\ref{fig:cache_BLIS}, which shows the effect of increasing
$k$ on the practical performance of the BLIS realization of the \gemm kernel.
The performance curve there points in the direction of a connection between the
utilization rate of the cache hierarchy and the performance of \gemm. In
Section~\ref{sec:experiments} we will make this connection explicit using hardware
counters.

\subsection{Model-driven cache configuration parameters}

The results in the table on the left of Figure~\ref{fig:cache_BLIS}
indicate that we can leverage
the BLIS micro-kernel $\mkdim{6}{8}$, yet 
attain a higher utilization of
the L1\textbar L2 cache levels by choosing larger values for $k_c$\textbar$m_c$. 
We next review how we can avoid a costly optimization search
of the CCP space by analytically selecting
the CCPs, depending on 
the specific cache hierarchy of the target architecture 
as well as the \gemm dimensions.

Once the dimensions of the micro-kernel are fixed, 
the analytical model in~\cite{BLIS4} 
takes into account the organization of the cache hierarchy 
(specifically,  number of levels, dimension, cache line size, and associativity),
in order to determine
optimal values for the 
CCPs $k_c^m\rightarrow m_c^m\rightarrow n_c^m$,
\textit{in that order}, 
which respectively favour a high usage of the L1$\rightarrow$L2$\rightarrow$L3 caches.
Concretely, for the NVIDIA ARM-based Carmel processor 
and $\mkdim{6}{8}$,
the original model in~\cite{BLIS4} selects the CCPs
\[
(m_c^m,n_c^m,k_c^m) = (672,480,340). 
\]
Note that these values are significantly larger that those fixed inside BLIS.
However, the original model selects
the optimal value $k_c^m$ independently of the actual dimension $k$ and, 
therefore, does not take into
account the restriction $k_c \le k$ when deciding the optimal value of the CCP 
$m_c$ for the next level. 

We next refine the 
analytical procedure in~\cite{BLIS4} that calculates
the CCPs to make them not only architecture-specific,
\textit{but also \gemm ``dimension-aware''.} 
For this purpose, in contrast with the original analytical model, our refined version 
considers the actual dimension of $k_c=\min(k,k_c^m)$ when choosing $m_c^m$;
and the same strategy applies for the next level, when we choose $n_c$ taking into
account the actual value of  $m_c$.

\begin{table}[t]
\normalsize
\caption{Theoretical cache occupation of $B_r$\textbar$A_c$ in the L1\textbar L2 caches for \gemm kernels with
$m=n=2000$ and $k\in\{[64,256],2000\}$, and $\mkdim{6}{8}$.
In those cases labeled in the first column with 
``\textsf{BLIS}'',  
$(m_c,n_c,k_c)=$ 
$(m_c^B,n_c^B,k_c^B)$ defined by BLIS; 
when the label is ``\textsf{MOD}'', 
$(m_c,n_c,k_c)=(m_c^M,n_c^M,k_c^M)$ are calculated using 
the refined analytical model.}
\label{tab:cache_microkernel}
\centering
\setlength{\tabcolsep}{4pt}
\begin{tabular}{|lr|rrrrr||rrr|rrr|}
\hline 
Cache   
& $k$ & $m_c$ & $n_c$ & $k_c$ & $m_r$ & $n_r$  & L1 &      &     & L2 &       & \\
params.
&     &         &          &        &       &        & KB & (\%) & Max & KB &  (\%) & Max 
     \\ \hline
\noalign{\vskip-2.3\tabcolsep \vskip-3\arrayrulewidth \vskip\doublerulesep}
\\ \hline 
\textsf{BLIS} &   64 &  120 & 2000 &   64 &   6 &   8 &  4.0 &  6.2 & -- &   60.0 &  2.9 & -- \\ 
\textsf{MOD} &   64 & 2000 &  512 &   64 &   6 &   8 &  4.0 &  6.2 & 50.0 & 1000.0 & 48.8 & 81.2 \\ 
\hline
\textsf{BLIS} &   96 &  120 & 2000 &   96 &   6 &   8 &  6.0 &  9.4 & -- &   90.0 &  4.4 & -- \\ 
\textsf{MOD} &   96 & 2000 &  336 &   96 &   6 &   8 &  6.0 &  9.4 & 50.0 & 1500.0 & 73.2 & 81.2 \\ 
\hline
\textsf{BLIS} &  128 &  120 & 2000 &  128 &   6 &   8 &  8.0 & 12.5 & -- &  120.0 &  5.9 & -- \\ 
\textsf{MOD} &  128 & 1792 &  256 &  128 &   6 &   8 &  8.0 & 12.5 & 50.0 & 1792.0 & 87.5 & 87.5 \\ 
\hline
\textsf{BLIS} &  160 &  120 & 2000 &  160 &   6 &   8 & 10.0 & 15.6 & -- &  150.0 &  7.3 & -- \\ 
\textsf{MOD} &  160 & 1424 &  400 &  160 &   6 &   8 & 10.0 & 15.6 & 50.0 & 1780.0 & 86.9 & 87.5 \\ 
\hline
\textsf{BLIS} &  192 &  120 & 2000 &  192 &   6 &   8 & 12.0 & 18.8 & -- &  180.0 &  8.8 & -- \\ 
\textsf{MOD} &  192 & 1184 &  336 &  192 &   6 &   8 & 12.0 & 18.8 & 50.0 & 1776.0 & 86.7 & 87.5 \\ 
\hline
\textsf{BLIS} &  224 &  120 & 2000 &  224 &   6 &   8 & 14.0 & 21.9 & -- &  210.0 & 10.3 & -- \\ 
\textsf{MOD} &  224 & 1024 &  432 &  224 &   6 &   8 & 14.0 & 21.9 & 50.0 & 1792.0 & 87.5 & 87.5 \\ 
\hline
\textsf{BLIS} &  256 &  120 & 2000 &  240 &   6 &   8 & 15.0 & 23.4 & -- &  225.0 & 11.0 & -- \\ 
\textsf{MOD} &  256 &  896 &  512 &  256 &   6 &   8 & 16.0 & 25.0 & 50.0 & 1792.0 & 87.5 & 87.5 \\ 
\hline
\textsf{BLIS} & 2000 &  120 & 2000 &  240 &   6 &   8 & 15.0 & 23.4 & -- &  225.0 & 11.0 & -- \\ 
\textsf{MOD} & 2000 &  672 &  480 &  341 &   6 &   8 & 21.3 & 33.3 & 50.0 & 1790.2 & 87.4 & 87.5 \\ 
\hline
\hline
\end{tabular}
\end{table}

Let us briefly discuss the practical differences between these
two model-driven alternatives using, for example, a \gemm kernel
of dimensions $m=n=2000$ and $k=224$, with $\mkdim{6}{8}$.
The original model chooses
$(m_c^m,n_c^m,k_c^m)$ as previously indicated, independently of the problem
dimensions, yet the actual 
$k_c=\min(k,k_c^m)=\min(224,340)=224$.
In comparison, taking into account the restriction on $k_c$, 
for this particular case the refined model selects
the alternative parameters 
\[
(m_c^M,n_c^M,k_c^M)= (1792,256,224).
\]
In other words, the fact that $k_c$ is 
in practice constrained by $k$ 
is taken into account in the refined model
in order to increase $m_c$, potentially yielding a higher utilization of the L2 cache.

Table~\ref{tab:cache_microkernel} reveals
the effect of the alternative CCPs on the
utilization of the cache hierarchy for $\mkdim{6}{8}$.
Specifically, the theoretical analysis in the table
reports the occupation of the L1\textbar L2 caches with $B_r$\textbar $A_c$ when using the
BLIS CCPs 
versus the values determined by the refined model.
For reference, the table also includes the maximum
utilization of the corresponding level according
to the refined model. 
Note that, in all cases, $(m_c,n_c,k_c) \le (m,n,k)$.
Thus, for example, 
$n_c^B = 3072$ for BLIS, but in practice
$n_c=\min(n,n_c^B) = 2000$.
For the same reason, when $k=128$, even though $k_c^B=240$ for BLIS,
we have that $k_c=\min(k,k_c^B)=128$.
The figures in the table clearly show
that using the same micro-kernel $\mkdim{6}{8}$ as in BLIS, 
we can significantly
improve the utilization of the L2 cache by choosing
$m_c=m_c^M$ instead of $m_c^B$. 
Although not interesting for the LU factorization,
when $k$ is large, it is also possible to attain a higher
occupation of the L1 cache by using the refined model CCPs instead of the CCPs set in BLIS.

\subsection{Alternative micro-kernels}

While current realizations of BLAS only integrate a single micro-kernel per architecture, 
\textit{we propose to deviate from this convention in order
to explore the effect of adopting micro-kernels of other dimensions (shapes).}
In theory, we could consider micro-kernels of many
dimensions but, in practice, it is convenient
to restrict the study to those cases which do not incur into register spilling~\cite{Dowd98} 
and
where at least one of  the dimensions $m_r,n_r$ is an integer multiple 
of the SIMD lane size (128 bits for the target ARMv8.2 architecture).
We remind here that the micro-kernel shape has other effects on performance as a
squarish micro-kernel improves
the flops-to-memops ratio. For
example, for $k_c=128$, the BLIS micro-kernel 
$\mkdim{6}{8}$ yields 6.5 flops-per-memop
versus 5.5 for $\mkdim{4}{10}$ and 5.7 for $\mkdim{4}{12}$.

\vspace*{1ex}
\noindent
\textbf{Extended theoretical analysis.}
Table~\ref{tab:cache_alternative_microkernels}
repeats the analysis of cache utilization for four micro-kernels
of different dimensions, when setting the CCPs via the refined model.
These results show that, for the L1 cache,
by abandoning the BLIS micro-kernel
in favour of $\mkdim{4}{10}$ or $\mkdim{4}{12}$,
we can attain mildly higher utilization rates.
When the alternative is
either $\mkdim{10}{4}$ or $\mkdim{12}{4}$ though,
the utilization rates diminish.
In contrast, for the L2 cache, the utilization rates do not vary, with 
the exception of a moderate decrease 
for $\mkdim{4}{10}, \mkdim{4}{12}$ when $k=64$ or 128.

\begin{table}[t]
\normalsize 
\caption{Theoretical cache occupation of $B_r$\textbar$A_c$ in the L1\textbar L2 caches for \gemm kernels with
$m=n=2000$ and $k\in\{[64,256],2000\}$, and different micro-kernels.}
\label{tab:cache_alternative_microkernels}
\centering
\setlength{\tabcolsep}{4pt}
\begin{tabular}{|lr|rrrrr||rrr|rrr|}
\hline 
Cache   
& $k$ & $m_c$ & $n_c$ & $k_c$ & $m_r$ & $n_r$  & L1 &      &     & L2 &       & \\
params.
&     &         &          &        &       &        & KB & (\%) & Max & KB &  (\%) & Max 
     \\ \hline
\noalign{\vskip-2.3\tabcolsep \vskip-3\arrayrulewidth \vskip\doublerulesep}
\\ \hline 
\textsf{MOD} &   64 & 2000 &  512 &   64 &   4 &  10 &  5.0 &  7.8 & 50.0 & 1000.0 & 48.8 & 75.0 \\ 
\textsf{MOD} &   64 & 2000 &  512 &   64 &   4 &  12 &  6.0 &  9.4 & 50.0 & 1000.0 & 48.8 & 75.0 \\ 
\textsf{MOD} &   64 & 2000 &  512 &   64 &  10 &   4 &  2.0 &  3.1 & 25.0 & 1000.0 & 48.8 & 87.5 \\ 
\textsf{MOD} &   64 & 2000 &  512 &   64 &  12 &   4 &  2.0 &  3.1 & 25.0 & 1000.0 & 48.8 & 87.5 \\ 
\hline
\textsf{MOD} &  128 & 1664 &  256 &  128 &   4 &  10 & 10.0 & 15.6 & 50.0 & 1664.0 & 81.2 & 81.2 \\ 
\textsf{MOD} &  128 & 1664 &  256 &  128 &   4 &  12 & 12.0 & 18.8 & 50.0 & 1664.0 & 81.2 & 81.2 \\ 
\textsf{MOD} &  128 & 1792 &  256 &  128 &  10 &   4 &  4.0 &  6.2 & 25.0 & 1792.0 & 87.5 & 87.5 \\ 
\textsf{MOD} &  128 & 1792 &  256 &  128 &  12 &   4 &  4.0 &  6.2 & 25.0 & 1792.0 & 87.5 & 87.5 \\ 
\hline
\textsf{MOD} &  192 & 1184 &  336 &  192 &   4 &  10 & 15.0 & 23.4 & 50.0 & 1776.0 & 86.7 & 87.5 \\ 
\textsf{MOD} &  192 & 1184 &  336 &  192 &   4 &  12 & 18.0 & 28.1 & 50.0 & 1776.0 & 86.7 & 87.5 \\ 
\textsf{MOD} &  192 & 1184 &  336 &  192 &  10 &   4 &  6.0 &  9.4 & 25.0 & 1776.0 & 86.7 & 87.5 \\ 
\textsf{MOD} &  192 & 1184 &  336 &  192 &  12 &   4 &  6.0 &  9.4 & 25.0 & 1776.0 & 86.7 & 87.5 \\ 
\hline
\textsf{MOD} &  256 &  896 &  512 &  256 &   4 &  10 & 20.0 & 31.2 & 50.0 & 1792.0 & 87.5 & 87.5 \\ 
\textsf{MOD} &  256 &  896 &  512 &  256 &   4 &  12 & 24.0 & 37.5 & 50.0 & 1792.0 & 87.5 & 87.5 \\ 
\textsf{MOD} &  256 &  896 &  512 &  256 &  10 &   4 &  8.0 & 12.5 & 25.0 & 1792.0 & 87.5 & 87.5 \\ 
\textsf{MOD} &  256 &  896 &  512 &  256 &  12 &   4 &  8.0 & 12.5 & 25.0 & 1792.0 & 87.5 & 87.5 \\ 
\hline
\end{tabular}
\end{table}

\vspace*{1ex}
\noindent
\textbf{Implementing the micro-kernels.}
The \gemm micro-kernels are in general encoded using assembly with
architecture-specific data prefetching instructions~\cite{HenP17}. 
From a high level perspective, 
the code for the \gemm micro-kernel is rather 
simple, consisting of a main loop which, at each iteration, 
loads one column of $A_r$ plus one row of $B_r$, 
to then update the full 
contents of the micro-tile $C_r$; see Figure~\ref{fig:blocked_GEMM} (top right). 
A possible implementation is illustrated for 
$\mkdim{6}{8}$ and $\mkdim{12}{4}$ in Figure~\ref{fig:micro-kernels}.
There, we employ C instructions and Neon vector intrinsics~\cite{neonweb} instead of 
assembly. 
Note that ARMv8 features 32 vector registers of 128 bits each, with capacity
for two FP64 numbers. 
$\mkdim{6}{8}$ employs 24 vector registers to store $C_r$,
3 for the column of $A_r$,
and 4 for the row of $B_r$, for a total of 31.
In comparison, 
$\mkdim{12}{4}$ employs 24 vector registers for $C_r$,
6 for $A_r$, and 2 for $B_r$, for a total of 32.

We have implemented our own version
of several micro-kernels, 
using both vector intrinsics and assembly.
By experimenting with them, we found that the ordering of the instructions
inside the loop 
has a relevant impact on performance due to the presence of
write-after-read (WAR) dependencies~\cite{HenP17} between instructions in consecutive iterations that refer to the entries
of $A$ in the vector registers, and which are not solved by the hardware
scheduler
at execution time. By ordering the instructions as shown in
Figure~\ref{fig:micro-kernels},
our $\mkdim{6}{8}$ matches the performance of the assembly-encoded one in BLIS.
For the micro-kernels that utilize the full set of vector registers, 
from the point of view of performance, it is important
to code in assembly instead of using vector instrinsics to enforce 
this explicit ordering of the instructions.

\begin{figure}[thb!]
\centering
\begin{tabular}{ccc}
\begin{minipage}[c]{0.42\textwidth}
\begin{lstlisting}[language=C,alsoletter={.},deletekeywords={.sum},morekeywords={matrix}]
bA = 0; bB = 0; mr = 6; nr = 8;
for ( k=0; k<kc; k++ ) {
  A0 = vload(&Aptr[bA]);
  A1 = vload(&Aptr[bA+2]);
  A2 = vload(&Aptr[bA+4]);

  B0 = vload(&Bptr[bB]);
  B1 = vload(&Bptr[bB+2]);
  B2 = vload(&Bptr[bB+4]);
  B3 = vload(&Bptr[bB+6]);

  vupdate(C00, A0, B0, 0);
  vupdate(C01, A0, B0, 1);
  vupdate(C02, A0, B1, 0);
  vupdate(C03, A0, B1, 1);
  vupdate(C04, A0, B2, 0);
  vupdate(C05, A0, B2, 1);
  vupdate(C06, A0, B3, 0);
  vupdate(C07, A0, B3, 1);

  vupdate(C10, A1, B0, 0);
  /* C11, C12, C13,...,C17    */
  /* omitted for brevity      */

  vupdate(C20, A2, B0, 0);
  /* C21, C22, C23,...,C27    */
  /* omitted for brevity      */

  bA = bA+mr;
  bB = bB+nr;
}
\end{lstlisting}

\end{minipage}
&~~~~~~&
\begin{minipage}[c]{0.42\textwidth}
\begin{lstlisting}[language=C,alsoletter={.},deletekeywords={.sum},morekeywords={matrix}]
bA = 0; bB = 0; mr = 12; nr = 4;
for ( k=0; k<kc; k++ ) {
  A0 = vload(&Aptr[bA]);
  A1 = vload(&Aptr[bA+2]);
  A2 = vload(&Aptr[bA+4]);
  A3 = vload(&Aptr[bA+6]);
  A4 = vload(&Aptr[bA+8]);
  A5 = vload(&Aptr[bA+10]);

  B0 = vload(&Bptr[bB]);
  B1 = vload(&Bptr[bB+2]);
 
  vupdate(C00, A0, B0, 0);
  vupdate(C01, A0, B0, 1);
  vupdate(C02, A0, B1, 0);
  vupdate(C03, A0, B1, 1);

  vupdate(C10, A1, B0, 0);
  vupdate(C11, A1, B0, 1);
  vupdate(C12, A1, B1, 1);
  vupdate(C13, A1, B1, 1);

  /* C20, C21, C22, C23       */
  /* C30, C31, C32, C33       */
  /* C40, C41, C42, C43       */
  /* C50, C51, C52, C53       */
  /* omitted for brevity      */

  bA = bA+mr;
  bB = bB+nr;
}

\end{lstlisting}
\end{minipage}
\end{tabular}
\begin{minipage}[c]{0.89\textwidth}
\begin{lstlisting}[language=C,alsoletter={.},deletekeywords={.sum},morekeywords={matrix}]
#define vload(vreg, mem)  vreg = vld1q_f64(mem)
#define vstore(mem, vreg) vst1q_f64(mem, vreg)
#define vupdate(vreg1, vreg2, vreg3, j) \
        vreg1 = vfmaq_laneq_f64(vreg1, vreg2, vreg3, j)

\end{lstlisting}
\end{minipage}
\caption{Main loop in
         micro-kernels $\mkdim{6}{8}$ (top left) and $\mkdim{12}{4}$ (top right),
         and corresponding macros (bottom) implemented
         using ARM Neon vector intrinsics.} 
\label{fig:micro-kernels}
\end{figure}

\section{Experimental Evaluation}
\label{sec:experiments}

In this section we evaluate the practical effect of
tuning the CCPs and selecting the appropriate micro-kernel on 
performance 
for the individual \gemm routines as well as the LU
factorization with partial pivoting. 
Because the ultimate objective of the optimizations is to accelerate the matrix
factorization, we keep the analysis of \gemm short by discussing
that operation using only a single core of the target platforms,
and illustrating the potential benefits in both sequential and parallel performance for the LU factorization.

\subsection{Setup}

\begin{figure}[t]
\centering 
\begin{tabular}{c}
\includegraphics[width=0.7\textwidth]{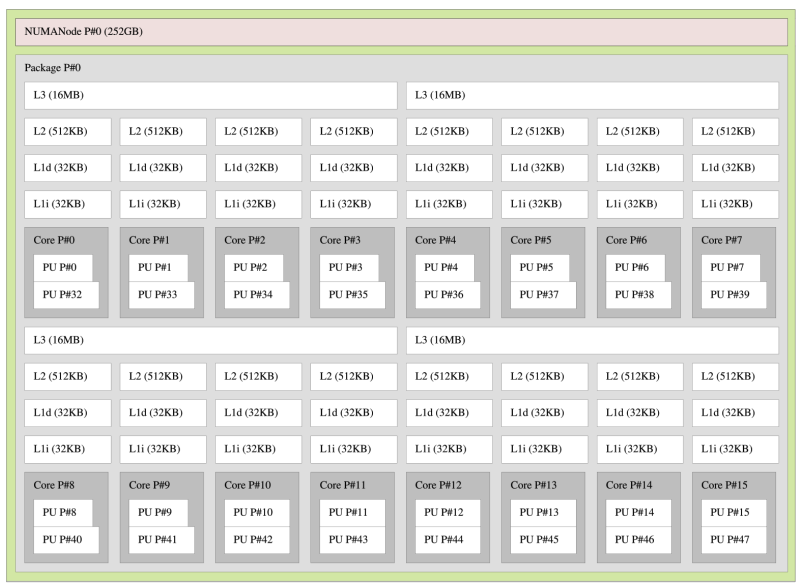}
\end{tabular}
\caption{Memory hierarchy of the AMD EPYC 7282 processor determined using \texttt{hwloc}.}
\label{fig:hwlock_amd_epyc}
\end{figure}

For the experiments, we leverage the NVIDIA Carmel processor already presented
in Section~\ref{sec:optimization}. In addition, we target
an x86-based platform, equipped
with a 16-core AMD EPYC 7282 processor furnished
with L1 (data), L2 and L3 cache memories with capacity for 32~KB per core, 512~KB per core and
16$\times 4$~MB shared by each group of four cores, respectively; see Figure~\ref{fig:hwlock_amd_epyc}.

With respect to the AMD EPYC 7282,
AMD's native implementation of BLAS, named AOCL, is basically BLIS
in disguise,\footnote{Specifically, the version of AOCL employed in our experiments corresponds to BLIS version 0.9.0, see \url{https://www.amd.com/en/developer/aocl/blis.html}.}
so we compare against the latter.
Prior to the experimental analysis, 
we make the following observations about the configuration of
BLIS on the AMD platform:
\begin{itemize}
\item For FP64 and this platform,
BLIS
encodes a micro-kernel 
$\mkdim{6}{8}$ that internally becomes
$\mkdim{8}{6}$ when $C$ is stored by columns.
\item BLIS features a \textsf{sup} variant
that integrates a collection of specialized micro-kernels when one or more of the \gemm dimensions
are small. For our particular platform, \textsf{sup} employs a non-packed micro-kernel
when $m,n$ are large but $k\le 256$. However, we observed a severe drop in performance 
in the \textsf{sup} implementation of \gemm
when the matrix operands are not aligned in memory (as is the case during the trailing update in the LU
factorization). 
To eliminate this effect, we configured BLIS to avoid the use of \textsf{sup}. 
\item The BLIS micro-kernel for the AMD processor integrates 
a sophisticated 
software prefetching scheme to reduce the memory access latency. 
In contrast, it is difficult 
to enhance our high-level micro-kernels based on AVX2 vector intrinsics 
with the equivalent type of instructions because that requires a 
thorough knowledge of the cache latencies, among other hardware
parameters.
To expose the effect of software prefetching on performance, 
in the experiments on the AMD server we evaluate the BLIS \gemm with and without this mechanism.
\item 
For FP64,
BLIS sets the CCPs as
$(m_c^B,n_c^B,k_c^B) = (72,2040,512)$ independently of the problem dimension. 
In comparison, for the 
$\mkdim{8}{6}$ and a \gemm with dimensions
$m=n=2000$, the refined analytical model for
example selects
$(m_c^M,n_c^M,k_c^M) = (768,2000,64)$ when $k=64$ and
$(192,2000,256)$ when $k=256$.
\end{itemize}

In both platforms, thread migration is prevented
via the appropriate Linux environment variable \texttt{OMP\_PROC\_BIND}. 
Furthermore, 
on the NVIDIA processor we fix the operation mode to \texttt{MAXN} which sets
the processor core's frequency to the maximum,
while
on the AMD socket we set the frequency to 2.3~GHz 
using the \texttt{cpufreq} tool. 

\subsection{NVIDIA Carmel}

\subsubsection{Matrix multiplication.}
We first explore the performance of three 
implementations of the \gemm routine, for brevity, limiting the analysis to a single core of the NVIDIA
Carmel processor: 

\begin{list}{}{}
\item R1) The realization in BLIS, with the BLIS $\mkdim{6}{8}$ micro-kernel and the CCPs statically fixed in BLIS. 
\item R2) Our own implementation of a BLIS-like \gemm with $\mkdim{6}{8}$.
\item R3) Our own implementation of a BLIS-like \gemm with $\mkdim{12}{4}$. 
\end{list}
\vspace*{-1ex}
In the last two cases, we employ micro-kernels encoded in assembly and
set the CCPs as dictated by the refined analytical model. 
In a separate experiment, we confirmed that the performance of
our implementation of \gemm with $\mkdim{6}{8}$, when setting the CCPs as in BLIS, 
matches that
of the BLIS \gemm routine. Therefore, the 
performance variations between R1 and R2 are strictly due to
the distinct values selected for the CCPs by the refined model.

\begin{figure}[tbh!]
\centering
\begin{tabular}{ccc}
\begin{minipage}[c]{0.64\textwidth}
\includegraphics[width=0.8\textwidth]{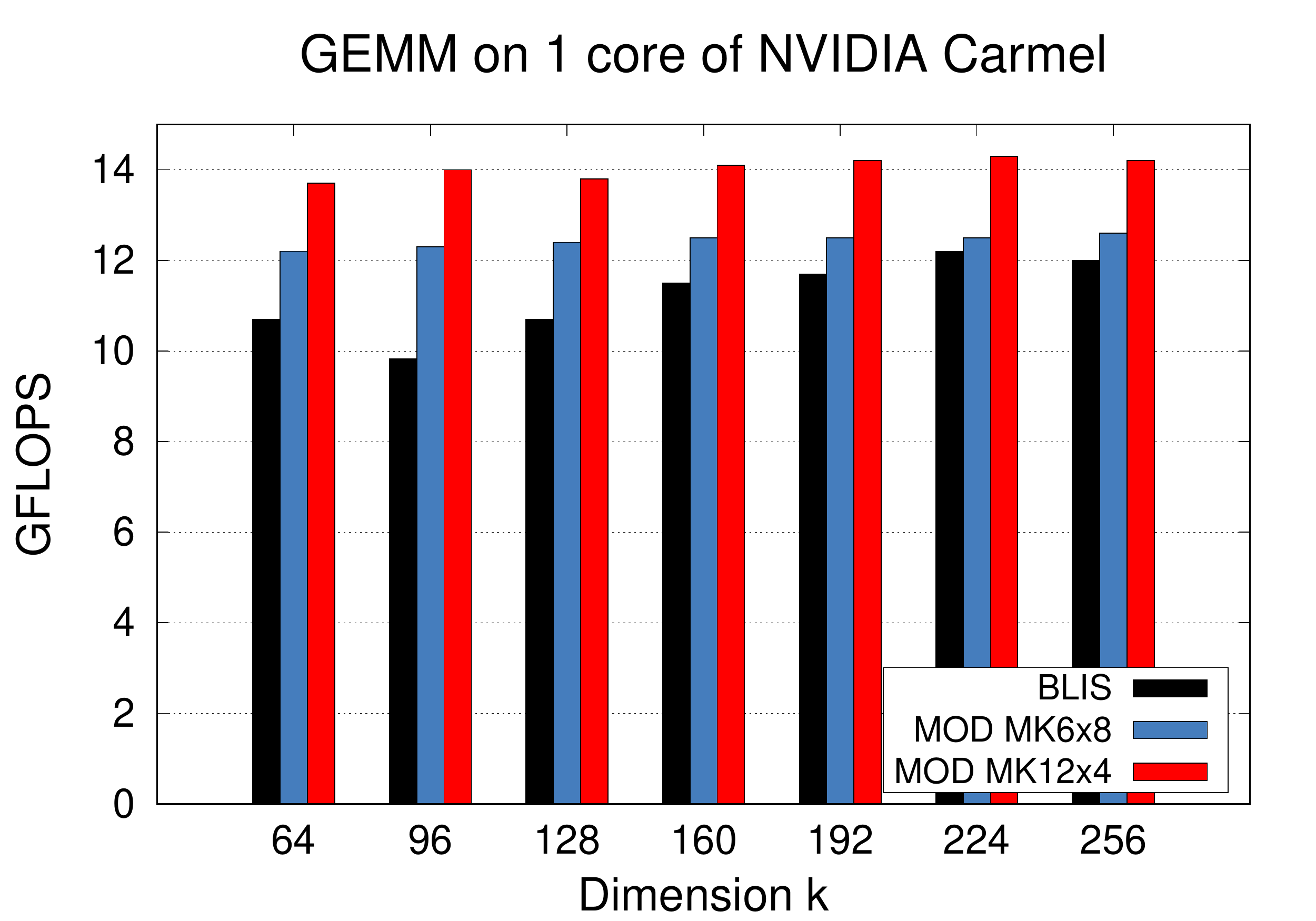}
\end{minipage}
& \hspace*{0.0cm} &
\begin{minipage}[c]{0.40\textwidth}
{\normalsize
\setlength{\tabcolsep}{4pt}
\begin{tabular}{|r||rr|}
\hline
    & \multicolumn{2}{c|}{Speed-up w.r.t. \textsf{BLIS}}   \\ \hline
$k$ & \textsf{MOD} & \textsf{MOD}  \\
    & $\mkdim{6}{8}$ & $\mkdim{12}{4}$
\\ \hline \hline

    64&     1.14&  1.28\\
    96&     1.25&  1.42\\
   128&     1.15&  1.28\\
   160&     1.08&  1.22\\
   192&     1.06&  1.21\\
   224&     1.02&  1.17\\
   256&     1.05&  1.18\\
     \hline
\end{tabular}
}
\end{minipage}
\end{tabular}
\caption{Evaluation of \gemm 
         with $m=n=2\,000$ and varying $k$ 
         using distinct CCPs and \gemm micro-kernels 
         on a single core of an NVIDIA Carmel processor.}
\label{fig:GEMM_perf}
\end{figure}

Figure~\ref{fig:GEMM_perf} reports the performance 
of the three \gemm routines for a problem
of dimension $m=n=2000$ and, because of its interest for matrix
decompositions in general and the LU factorization in particular, 
much smaller values for $k$, in the range $[64,256]$.
This initial experiment, using a single NVIDIA Carmel core, confirms the practical benefits
of 
taking into account both the target architecture and \gemm dimensions
when setting the CCPs  and choosing the micro-kernel following the theoretical
analysis in Section~\ref{sec:optimization}.
The gain due to the former factor (CCPs) is especially visible when $k$ is small,
yielding performance improvements of up to 42\% for $k=96$, and 
also overperforming BLIS for its best case (when $k=224$) with a performance
improvement of 18\%. 
The second factor (specialized micro-kernel) helps to attain a high 
arithmetic throughput, almost independent of the value of $k$.

Although not reported in the results, we also evaluated other micro-kernels
(e.g., 
$\mkdim{4}{10}$,
$\mkdim{4}{12}$,
$\mkdim{10}{4}$,\ldots). In general though, 
$\mkdim{12}{4}$ consistently produced the highest arithmetic throughput.
Together with Table~\ref{tab:cache_alternative_microkernels},
this points in the direction that, on this particular setting,
augmenting the utilization of the L1 cache  is not relevant but
the key is maximizing the usage of the L2 cache.
Also, even though 
$\mkdim{12}{4}$ is a very ``non-squarish'' micro-kernel,
its flop-per-memop ratio 
seems to be sufficiently large to hide 
the memory accesses from it.
In relation with this, an analysis of the codes for 
$\mkdim{12}{4}$,
$\mkdim{4}{12}$ revealed a larger number of WAR dependencies between consecutive
iterations in the latter, which we believe is the cause for its lower performance. 

\subsubsection{LU factorization.}
We next assess whether the gains due to the adoption of
tuned CCPs and specialized micro-kernels
for the \gemm carry over
to the LU factorization with partial pivoting on the NVIDIA Carmel processor. 
For this purpose, we
consider the decomposition of a square matrix of
order 
$s=10,000$, and set the algorithmic 
block size $b$ in the range $[64,256]$.
This yields a sequence of trailing updates, each involving a 
\gemm with decreasing dimensions $m=n\leq s$ as the factorization 
progresses, but constant $k=b$.
For the parallel execution, we follow the insights in~\cite{BLIS2} to parallelize loop 
\textsf{G4} for both the BLIS \gemm and our own implementations. 
Nonetheless, we also evaluated parallel versions that extracted parallelism
from loop \textsf{G3} but, as expected from the study 
in~\cite{BLIS2} for platforms with shared L2 cache, 
those codes consistently delivered lower performance for both BLIS and our own routines than the counterparts
that extract parallelism from loop \textsf{G4}.
Therefore, we omit them from the following evaluation.

\begin{figure}[tbh!]
\normalsize   
\centering
\begin{tabular}{ccc}
\begin{minipage}[c]{0.48\textwidth}
\includegraphics[width=0.8\textwidth,angle=-90]{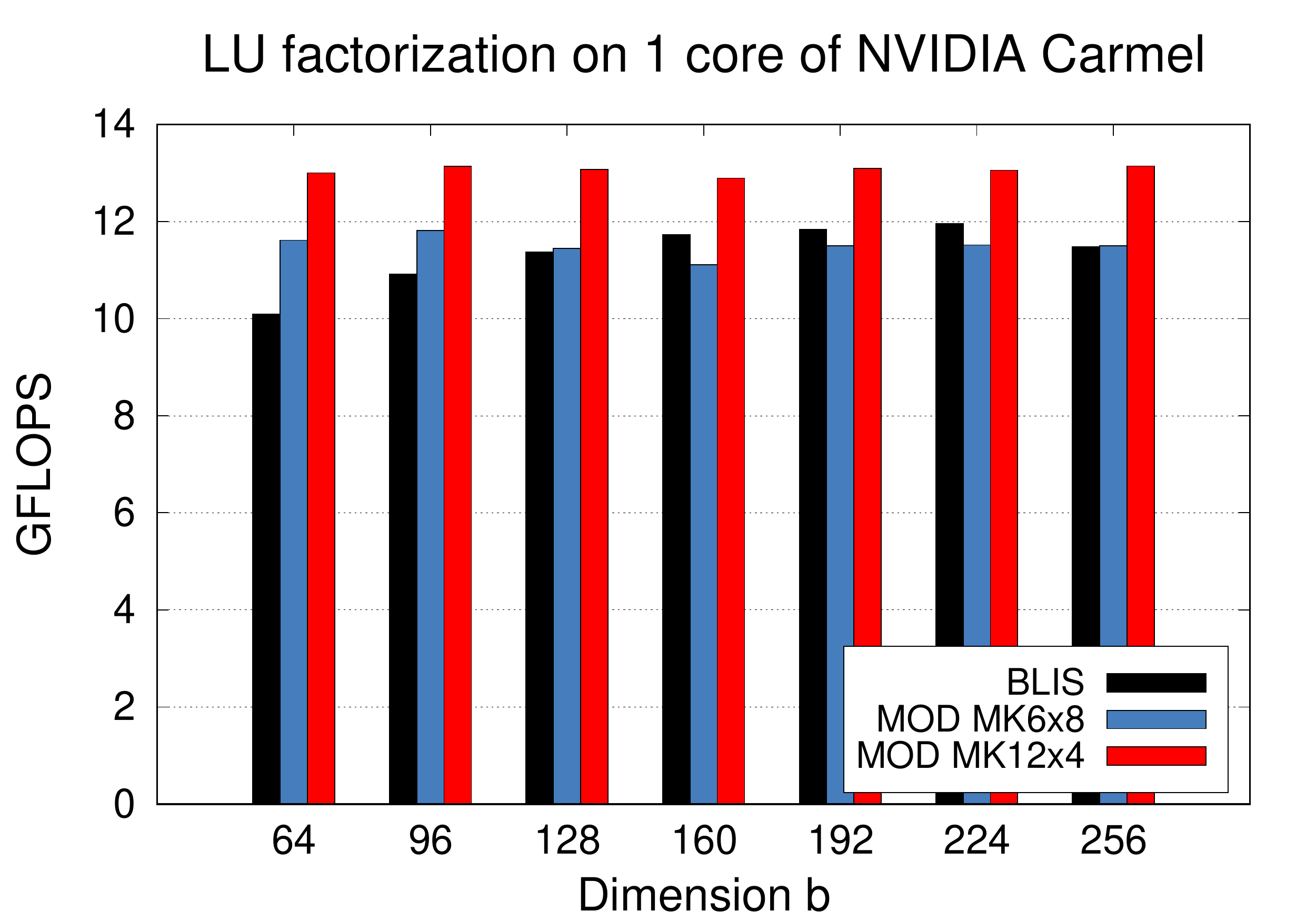} 
\end{minipage}
& \hspace*{0.2cm} &
\begin{minipage}[c]{0.40\textwidth}
{\normalsize 
\setlength{\tabcolsep}{4pt}
\begin{tabular}{|r||rr|}
\hline
    & \multicolumn{2}{c|}{Speed-up w.r.t. \textsf{BLIS}}   \\ \hline
$k$ & \textsf{MOD} & \textsf{MOD}  \\
    & $\mkdim{6}{8}$ & $\mkdim{12}{4}$ 
\\ \hline \hline
    64&    1.15&  1.28\\
    96&     1.08&  1.21\\
   128&     1.00&  1.14\\
   160&     0.94&  1.09\\
   192&     0.97&  1.12\\
   224&     0.97&  1.09\\
   256&     0.98&  1.12\\
\hline
\end{tabular}
}
\end{minipage}
\\
\begin{minipage}[c]{0.48\textwidth}
\includegraphics[width=0.8\textwidth,angle=-90]{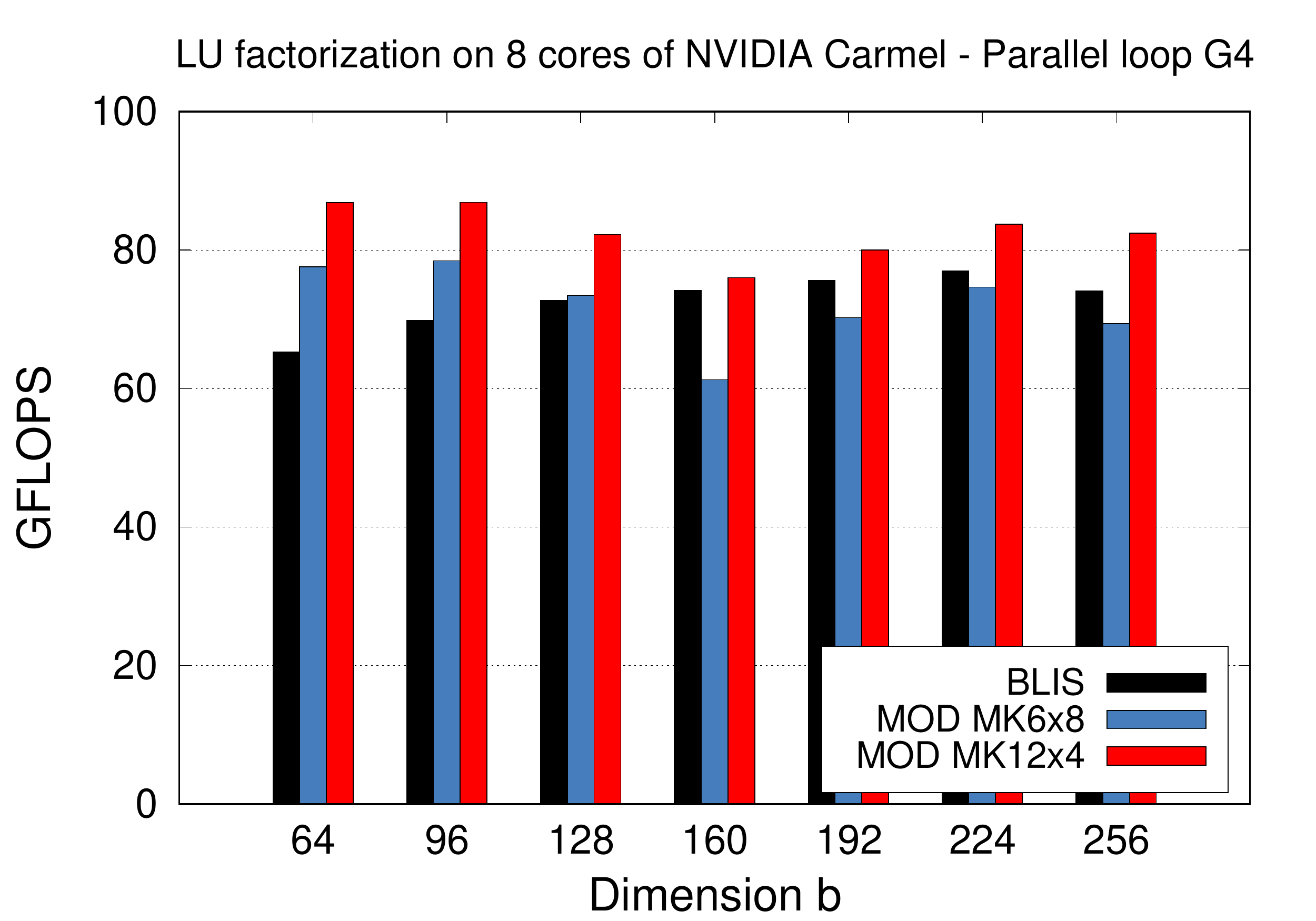} 
\end{minipage}
& \hspace*{2.2cm} &
\begin{minipage}[c]{0.40\textwidth}
{\normalsize
\setlength{\tabcolsep}{4pt}
\begin{tabular}{|r||rr|}
\hline
    & \multicolumn{2}{c|}{Speed-up w.r.t. \textsf{BLIS}}   \\ \hline
$k$ & \textsf{MOD} & \textsf{MOD}  \\
    & $\mkdim{6}{8}$ & $\mkdim{12}{4}$
\\ \hline \hline
    64&     1.18&     1.33\\
    96&     1.12&     1.24\\
   128&     1.00&     1.13\\
   160&     0.82&     1.02\\
   192&     0.92&     1.05\\
   224&     0.97&     1.08\\
   256&     0.93&     1.11\\
\hline
\end{tabular}
}
\end{minipage}
\end{tabular}
\caption{Performance of the LU factorization for a square matrix of order $s=10,000$
         and varying $b$, 
         using distinct CCPs and \gemm micro-kernels,
         on a single core (top)
         and the full socket parallelizing loop \textsf{G4} (bottom)
         of an NVIDIA Carmel processor.} 
\label{fig:LU_carmel}         
\end{figure}

The results in Figure~\ref{fig:GEMM_perf} illustrated that
the performance of the BLIS \gemm routine grows with $k$. 
However, the 
experiments with the LU factorization in Figure~\ref{fig:LU_carmel} depict a different
scenario: As $b$ is increased, the panel factorization 
considerably augments its relative cost, rapidly becoming a performance
bottleneck. This is especially the case when the trailing
update is performed in parallel, using all eight cores
of the NVIDIA Carmel processor.
In contrast, by adopting a specialized micro-kernel
such as $\mkdim{12}{4}$,
the performance of \gemm does not suffer such a large performance penalty in 
case $k$ is small, which allows to adopt a smaller algorithmic block size and helps in restricting the amount
of flops carried out in terms of the mostly-sequential panel
factorization. Ultimately, this factor determines the higher performance of the \gemm routine that
integrates 
$\mkdim{12}{4}$ with the CCPs selected by the refined analytical model.
All in all, the maximum performance improvements from a proper selection of CCPs and micro-kernel shapes
in GEMM when applied to the LU factorization vary between 28\% for the sequential version and 33\% for 
its parallel counterpart.

\subsection{AMD EPYC 7282}

\subsubsection{Matrix multiplication.}
We now analyze 
the performance of  four
\gemm routines on a single core of the AMD server:

\begin{list}{}{}
\item R1) \textsf{BLIS} without (software) prefetching.
\item R2) \textsf{BLIS} with (software) prefetching.
\item R3) Our own implementation of a BLIS-like \gemm with  $\mkdim{6}{8}$.
\item R4) Our own implementation of a BLIS-like \gemm with $\mkdim{8}{6}$.
\end{list}
For the latter two, the CCPs are set using the refined analytical model,
the corresponding micro-kernels are encoded using vector intrinsics 
and, as discussed at the beginning of this section, they
do not include any type of software prefetching.
Our micro-kernel $\mkdim{8}{6}$ has the
same dimensions as that in BLIS but it is 
implemented using vector intrinsics that
are compiled into the same SIMD assembly instructions that are present
in the BLIS micro-kernel. The only differences between BLIS
and our implementation of \gemm, when the software prefetching mechanism is inactive, lies
in the optimization of the packing routines and
the selection of the CCPs. In an independent experiment, we
could determine that the cost of packing the operands
is in general minor and, therefore, the performance differences 
between R1 and R3 originate from the distinct CCPs. 

We also implemented other micro-kernels (e.g., $\mkdim{10}{4}$, $\mkdim{12}{4}$, etc.) which, 
according to the model, should make a more efficient use of the
L1 cache. 
However, 
when the CCPs are properly set, 
due to the small capacity of the L2 cache on the AMD processor,
all the 
micro-kernels basically utilize the same space in that memory level.
Furthemore, as these alternative micro-kernels
are ``less square'', they
present a lower flops-memops ratio compared with
$\mkdim{6}{8}$,
$\mkdim{8}{6}$, which explains 
why we did not appreciate any benefit from them. 
In consequence, for simplicity, we exclude them from the presentation
of the following results.

\begin{figure}[thb!]
\begin{center}
\normalsize   
\centering
\begin{tabular}{ccc}
\begin{minipage}[c]{0.68\textwidth}
\includegraphics[width=0.8\textwidth]{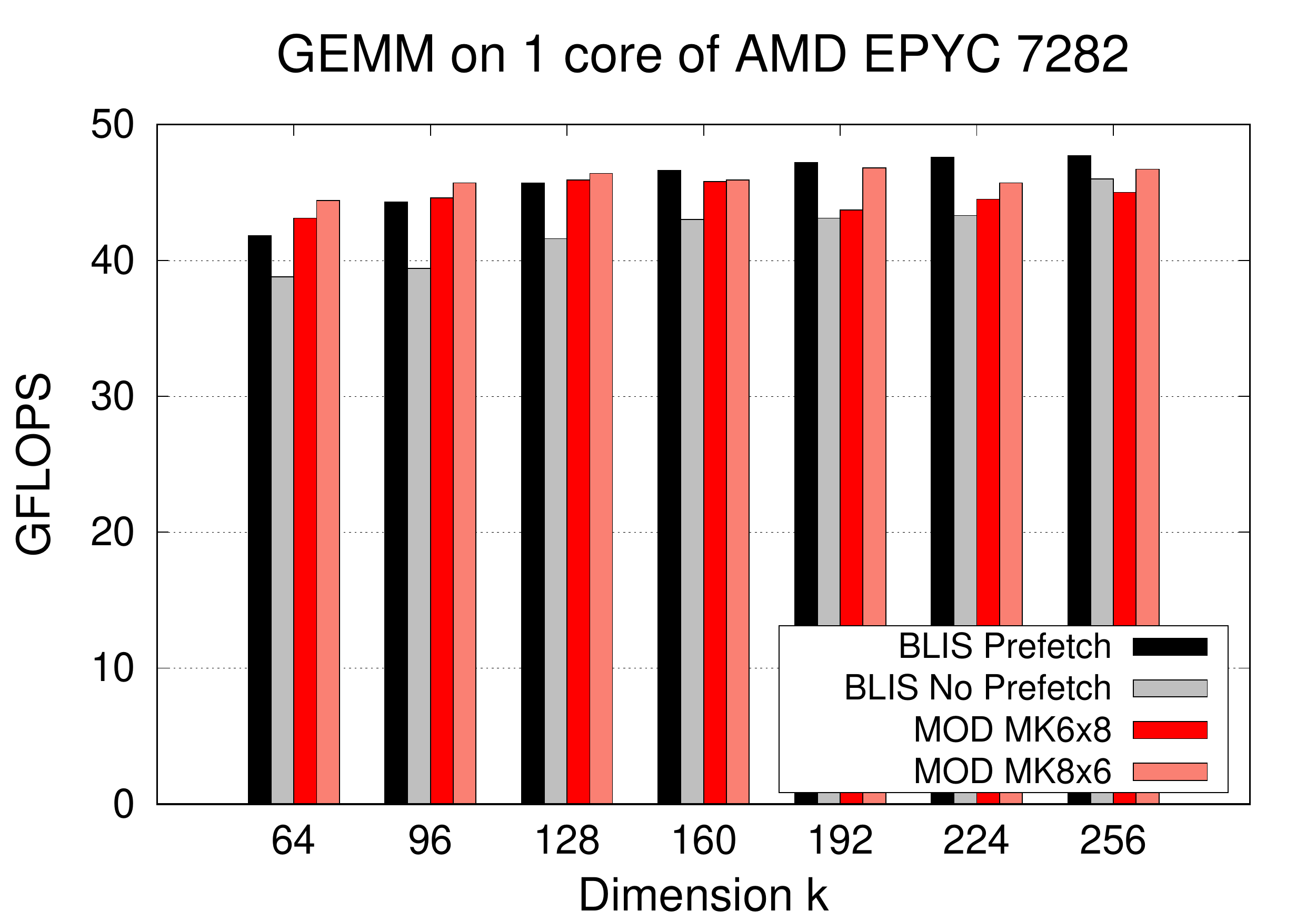}\\
\end{minipage}
& \hspace*{0.0cm} &
\begin{minipage}[c]{0.30\textwidth}
{\normalsize
\setlength{\tabcolsep}{4pt}
\begin{tabular}{|r||rrr|}
\hline
    & \multicolumn{3}{c|}{Speed-up w.r.t. \textsf{BLIS}}   \\ 
    & \multicolumn{3}{c|}{No Prefetch}   \\ \hline
$k$ & \textsf{BLIS} & \textsf{MOD} & \textsf{MOD}  \\
    &               & $\mkdim{6}{8}$ & $\mkdim{8}{6}$
\\ \hline \hline
     64&     1.08 & 1.11 & 1.14 \\
     96&     1.12 & 1.13 & 1.16 \\
    128&     1.10 & 1.10 & 1.12 \\
    160&     1.08 & 1.07 & 1.07 \\
    192&     1.10 & 1.01 & 1.09 \\
    224&     1.10 & 1.03 & 1.06 \\
    256&     1.04 & 0.98 & 1.02 \\
\hline
\end{tabular}
}
\end{minipage}
\\
\begin{minipage}[c]{0.68\textwidth}
\includegraphics[width=0.8\textwidth]{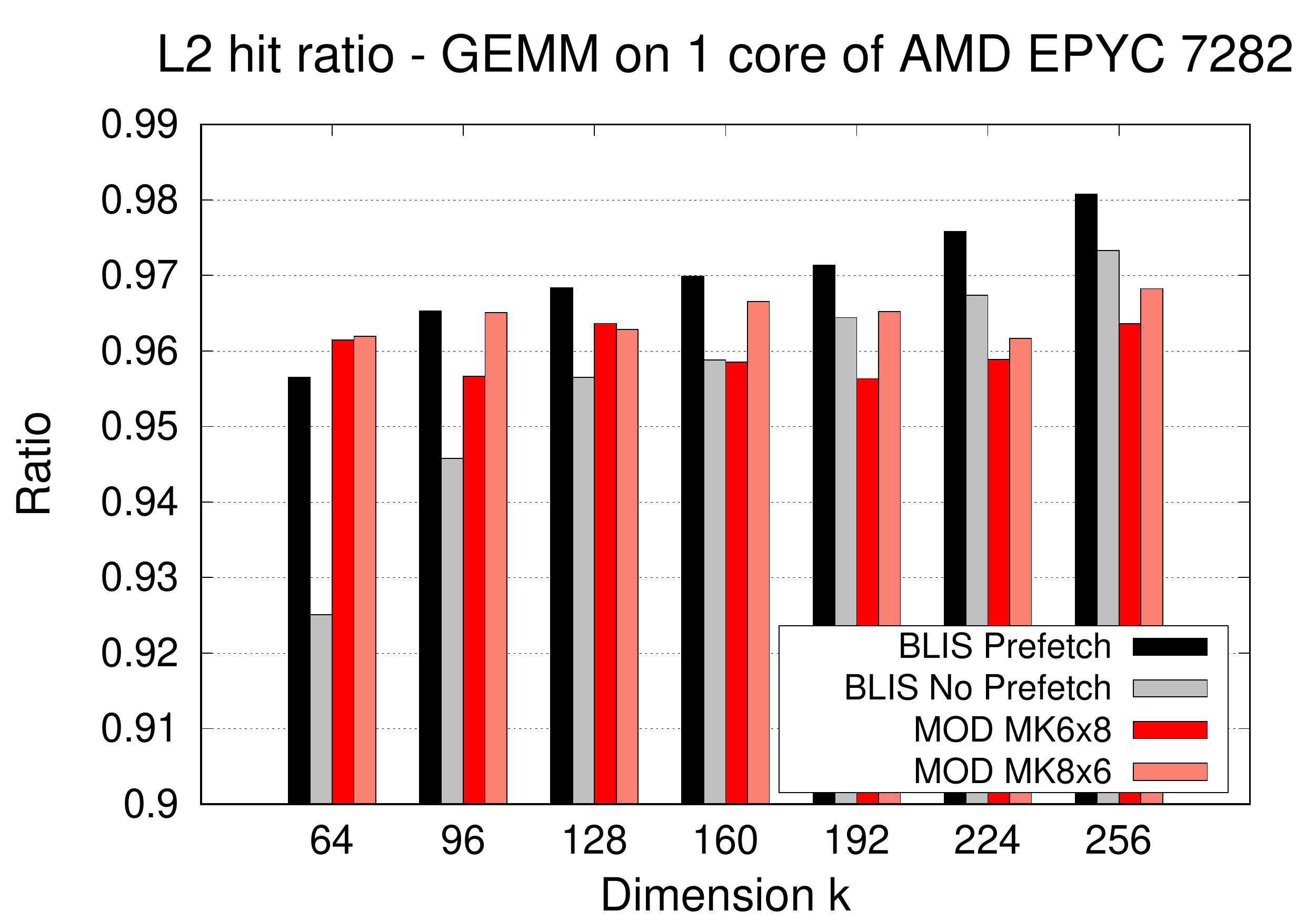}
\end{minipage}
\end{tabular}

\end{center}
\caption{Performance (top) and L2 hit ratio (bottom) of \gemm 
         with $m=n=2\,000$ and varying $k$, 
         using distinct CCPs and \gemm micro-kernels,
         on a single core of an AMD EPYC 7282 processor.}
\label{fig:AMD_GEMM_EPYC} 
\end{figure}

The plot and table in the top part of Figure~\ref{fig:AMD_GEMM_EPYC} 
report the evaluation of the \gemm routines for a problem
of dimension $m=n=2000$ and $k$ in the range $[64,256]$.
From the results we can make the following observations:
\begin{itemize}
\item When the software prefetching scheme is inactive,
      a proper selection of the CCPs, which takes into account 
      the actual value $\min(k,k_c)$ to set $m_c$ for the 
      L2 cache, is key to attain 
      high performance when $k$ is small. In addition, 
      as predicted by the model, when $k$ grows the differences are blurred.
\item The $\mkdim{8}{6}$ micro-kernel provides an extra yet small increase in performance with respect to
      $\mkdim{6}{8}$.
\item The software prefetching scheme is critical in this platform, helping to hide the memory access latency and
      in practice attaining a similar effect to that of a proper selection of the CCPs.
\item The observed performance gains are more modest than those observed for the Carmel processor. For AMD, the improvement
       ranges from 14\% ($k=64$) to 2\% ($k=256$).
\end{itemize}
To close this study of \gemm on the AMD EPYC 7282 processor, the bottom plot
in Figure~\ref{fig:AMD_GEMM_EPYC} shows the L2 cache hit ratio for the evaluated options,
demonstrating that the observed performance differences are coherent with the information exposed by this hardware
counter.

\subsubsection{LU factorization.}
For the AMD server, we finally consider
the LU decomposition with partial pivoting of a
square matrix of order $s = 10,000$ while
setting $b$ in the range $[64, 256]$.
For the parallel execution, we consider two options that differ in the loop that is parallelized: either loop \textsf{G3} or loop \textsf{G4}.

\begin{figure}[thb!]
\begin{center}
\begin{tabular}{ccc}
\begin{minipage}[c]{0.64\textwidth}
\includegraphics[width=0.8\textwidth]{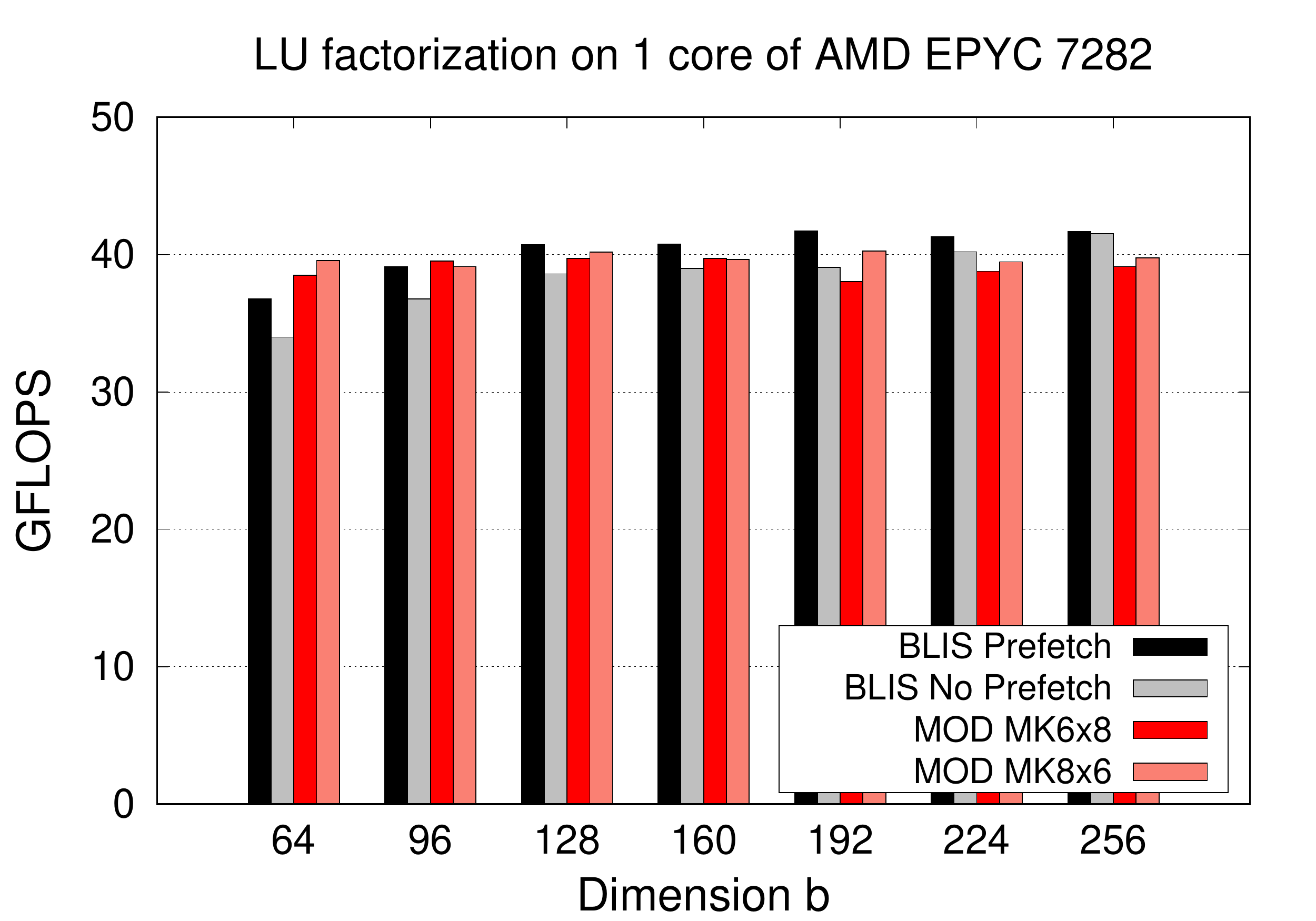}
\end{minipage}
& \hspace*{0.0cm} &
\begin{minipage}[c]{0.30\textwidth}
{\normalsize
\setlength{\tabcolsep}{4pt}
\begin{tabular}{|r||rrr|}
\hline
    & \multicolumn{3}{c|}{Speed-up w.r.t. \textsf{BLIS}}   \\ 
    & \multicolumn{3}{c|}{No Prefetch}   \\ \hline
$k$ & \textsf{BLIS} & \textsf{MOD} & \textsf{MOD}  \\
    &               & $\mkdim{6}{8}$ & $\mkdim{8}{6}$
\\ \hline \hline
     64&  1.08 & 1.13 & 1.16\\
     96&  1.06 & 1.08 & 1.06\\
    128&  1.05 & 1.03 & 1.04\\
    160&  1.04 & 1.02 & 1.02\\
    192&  1.07 & 0.97 & 1.03\\
    224&  1.03 & 0.96 & 0.98\\
    256&  1.00 & 0.94 & 0.96\\

\hline
\end{tabular}
}
\end{minipage}
\\
\begin{minipage}[c]{0.64\textwidth}
\includegraphics[width=0.8\textwidth]{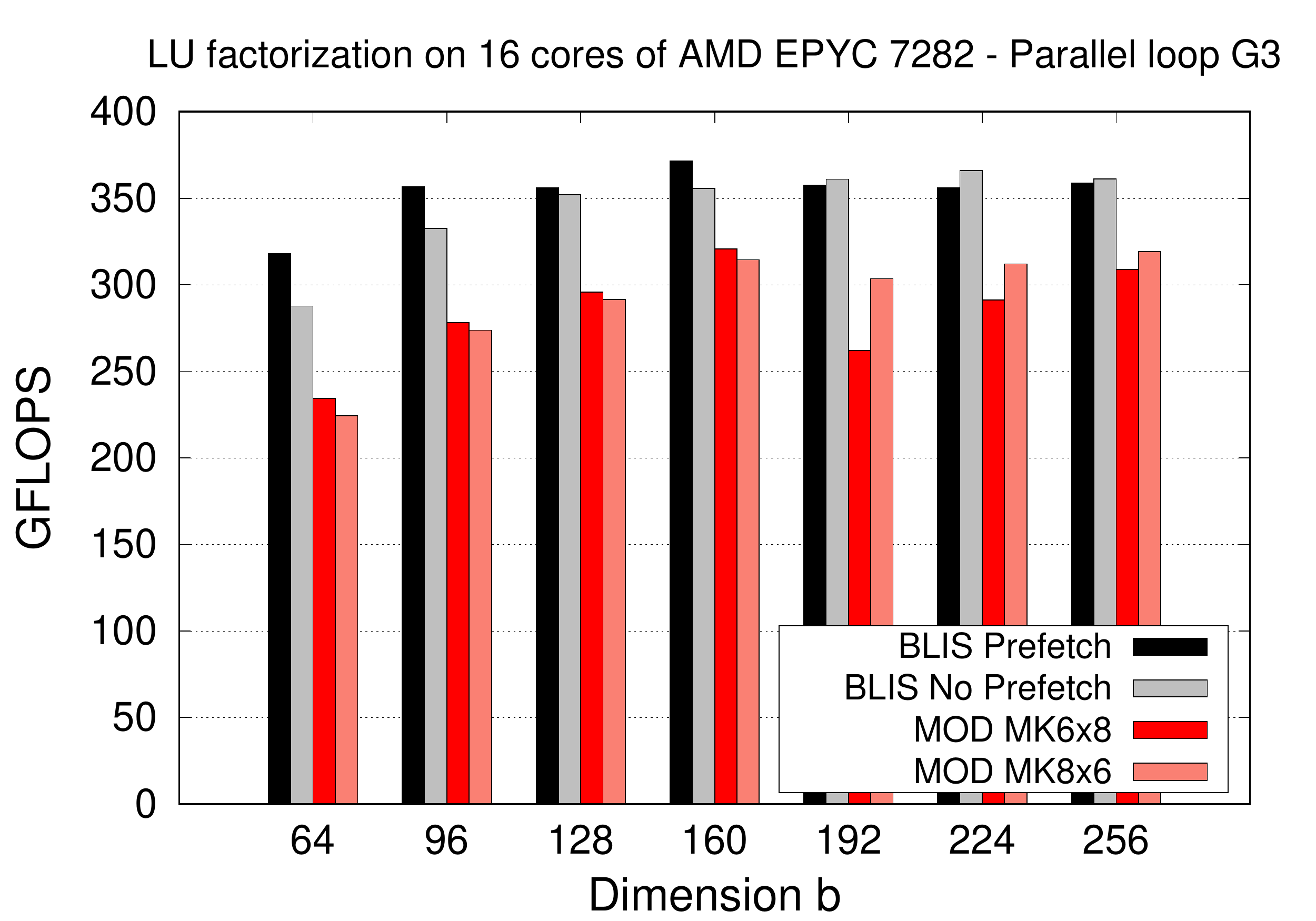}
\end{minipage}
& \hspace*{0.2cm} &
\begin{minipage}[c]{0.30\textwidth}
{\normalsize
\setlength{\tabcolsep}{4pt}
\begin{tabular}{|r||rrr|}
\hline
    & \multicolumn{3}{c|}{Speed-up w.r.t. \textsf{BLIS}}   \\ 
    & \multicolumn{3}{c|}{No Prefetch}   \\ \hline
$k$ & \textsf{BLIS} & \textsf{MOD} & \textsf{MOD}  \\
    &               & $\mkdim{6}{8}$ & $\mkdim{8}{6}$
\\ \hline \hline
     64&     1.11 & 0.81 & 0.78\\
     96&     1.07 & 0.84 & 0.82\\
    128&     1.01 & 0.84 & 0.83\\
    160&     1.04 & 0.90 & 0.88\\
    192&     0.99 & 0.73 & 0.84\\
    224&     0.97 & 0.80 & 0.85\\
    256&     0.99 & 0.86 & 0.88\\

\hline
\end{tabular}
}
\end{minipage}
\\
\begin{minipage}[c]{0.64\textwidth}
\includegraphics[width=0.8\textwidth]{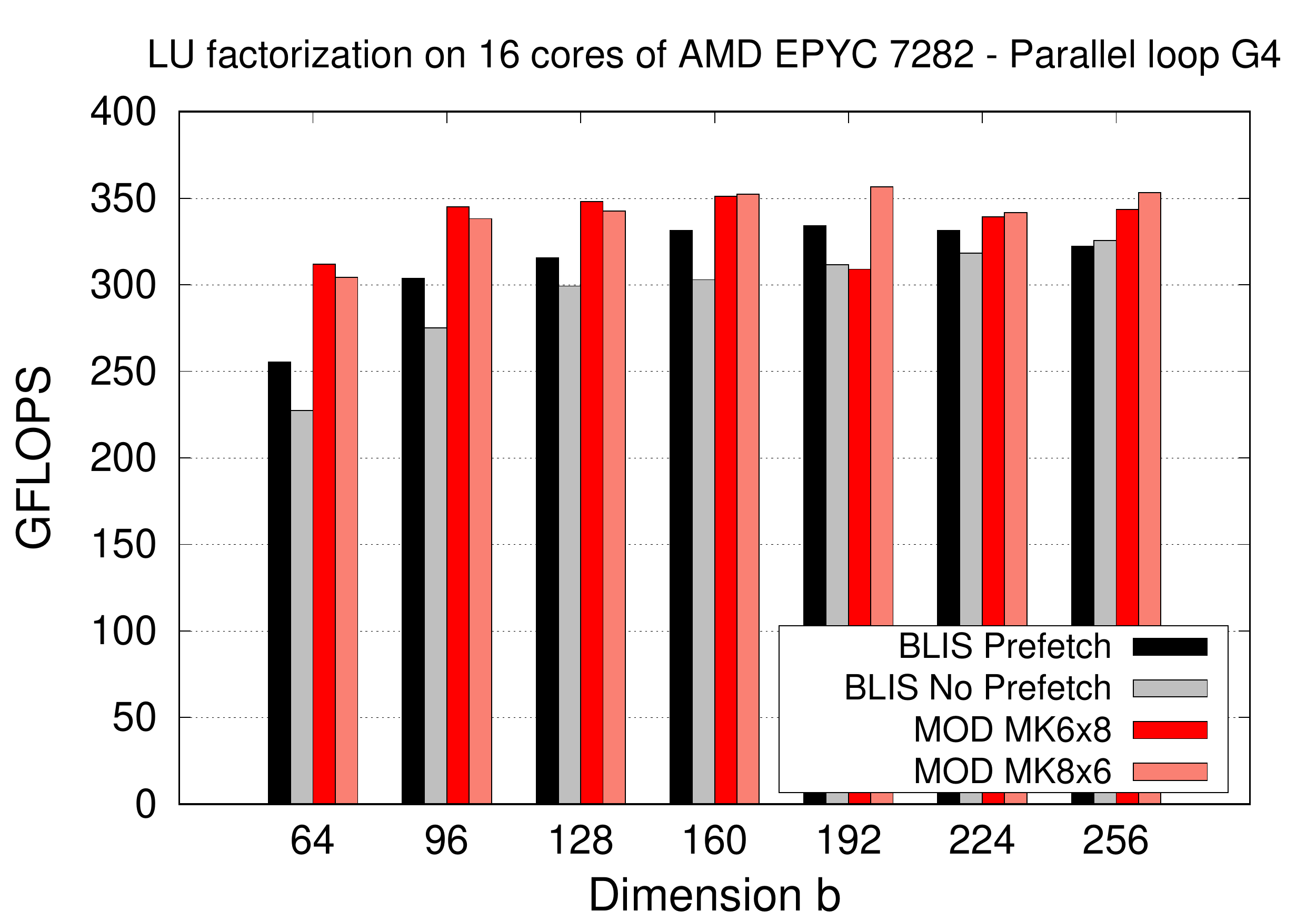}
\end{minipage}
& \hspace*{0.2cm} &
\begin{minipage}[c]{0.30\textwidth}
{\normalsize
\setlength{\tabcolsep}{4pt}
\begin{tabular}{|r||rrr|}
\hline
    & \multicolumn{3}{c|}{Speed-up w.r.t. \textsf{BLIS}}   \\ 
    & \multicolumn{3}{c|}{No Prefetch}   \\ \hline
$k$ & \textsf{BLIS} & \textsf{MOD} & \textsf{MOD}  \\
    &               & $\mkdim{6}{8}$ & $\mkdim{8}{6}$
\\ \hline \hline
     64&     1.12 & 1.37 & 1.34\\
     96&     1.10 & 1.25 & 1.23\\
    128&     1.05 & 1.16 & 1.15\\
    160&     1.09 & 1.16 & 1.16\\
    192&     1.07 & 0.99 & 1.14\\
    224&     1.04 & 1.07 & 1.07\\
    256&     0.99 & 1.05 & 1.08\\

\hline
\end{tabular}
}
\end{minipage}
\end{tabular}
\end{center}
\caption{Performance of the LU factorization for a square matrix of order $s=10,000$
         and varying $b$, 
         using distinct CCPs and \gemm micro-kernels,
         on a single core (top),
         the full socket parallelizing loop \textsf{G3} (middle)
         and the full socket parallelizing loop \textsf{G4} (bottom)
         of an ADM EPYC 7282 processor.}
\label{fig:LU_GEMM_LOOP_AMD} 
\end{figure}

From the results in 
Figure~\ref{fig:LU_GEMM_LOOP_AMD}, we 
highlight the following aspects:
\begin{itemize}
\item For the single-thread execution (top plot), 
      when the software prefetching mechanism is inactive and $k$ is small, it becomes critical
      to make a proper selection of the CCPs. However, as the capacity of the L2 cache is considerably more
      reduced in the AMD server than in the NVIDIA one, 
      a much smaller value of $k$ already ensures a close-to-optimal usage of the L2 cache in
      the former processor (for example, the performance improvement for $k=64$ compared with BLIS is 16\%; at that
      point the performance attained by our solution already attains the best performance for the LU factorization).
\item For the multi-threaded execution, the two parallelization options, targeting loops 
      \textsf{G3} or \textsf{G4}, offer quite different results.      
      \begin{itemize}
      \item When the parallelism is extracted from loop \textsf{G3} (middle plot), BLIS outperforms our implementations by
      a significant margin. The reason for this behavior is the lack of sufficient
      parallelism and/or an unbalanced workload distribution, which blurs the effect of the distinct utilization
      of the cache hierarchy. In detail, 
      BLIS statically sets $m_c^B=72$ while the refined analytical model 
      selects larger values for this parameter,
      in the range between $m_c^M=192$ for $b=256$ and $m_c^M=768$ for $b=64$.
      As a result, for example, with $b=192$, $m_c^M=384$,
      and even in the first iteration of the LU
      factorization, when $s=m=n=10,000$, there is little parallelism to ``feed'' all 16 threads as
      $m/m_c^M/16=10,000/384/16=1.62$ iterations per thread. In addition, this reveals the conditions for 
      significant workload imbalance.
      \item The situation is very different when the parallelization target is loop \textsf{G4} (bottom plot), with
      the results being more similar to those observed on the NVIDIA platform. In contrast with the
      parallel loop \textsf{G3}, the dimension that is parallelized is $n_c$ but the ``distribution'' size 
      is much smaller, $n_r$, which offers enough parallelism for 16 threads. In this scenario, the better
      utilization of the memory hierarchy helps to improve the arithmetic throughput rate in a visible way.
      \end{itemize}
\end{itemize}

\section{Concluding Remarks}
\label{sec:remarks}

We have presented a novel solution to improve the performance on modern multi-core architectures
of LAPACK-level blocked routines 
based on highly tuned BLAS implementations.
In order to improve the performance of \gemm operating with rectangular matrices,
1) our approach is armed with a refined analytical model for cache configuration parameters 
that takes into account the problem dimension as an additional input; 
and 2) unlike modern BLAS libraries, integrates and select the best option
among several micro-kernels, depending on the specific operand operand dimensions.
As a result,  
higher-level blocked algorithms (e.g., matrix factorizations) 
that make use of this primitive, show remarkable performance gains mainly due to better cache utilization, 
in particular the L2 cache, which has proven to be the key to performance improvement. 

This approach has been validated for both the \gemm kernel and the LU factorization, with sequential and parallel experiments, 
on two state-of-the-art multi-core architectures (ARM --NVIDIA Carmel-- and x86 --AMD EPYC--) with distinct vector lengths and cache systems. For the sequential implementations, the performance gains range from 28\% for the NVIDIA Carmel to
16\% for the AMD EPYC for the LU factorization; 
in this case, the larger L2 cache in the Carmel processor provides more performance
improvement space for our strategy. For the parallel implementations of the same operation, the performance gains
reach up to 33\% for the NVIDIA Carmel.
In the case of the AMD EPYC processor, the experimental results show that
the balance between cache utilization and workload distribution can limit performance gains on multi-core architectures
with many cores; however, our approach still offers significant performance gains (up to 34\%) in scenarios with proper workload distribution across threads.

The experimental results support the main message of the paper: on architectures equipped with hierarchical caches, the historically rigid boundary between the BLAS and the higher-level libraries that use its routines (e.g., LAPACK) should be permeated,
resulting in the co-design of the DLA software stack that yields a more efficient utilization of the cache hierarchy. In practice, this requires that the optimized BLAS libraries take into account not only the micro-architectural features of the underlying processor, but also the specific dimensions and shapes of the operands. From this perspective, tuned BLAS realizations should no longer be monolithic 
in the selection of cache configuration parameters and micro-kernels, but rather flexibly adapt their selection to each specific use case.

\begin{acks}
This work was supported by grants
PID2020-113656RB-C22 and  PID2021-126576NB-I00 of MCIN/AEI/10.13039/501100011033, 
and by {\em ``ERDF A way of making Europe''}.

This project has received funding from the European High-Performance Computing Joint Undertaking (JU) under grant agreement No
955558. The JU receives support from the European Union’s Horizon 2020 research and innovation programme, 
and Spain, Germany, France, Italy, Poland, Switzerland, Norway.

Héctor Martínez is a postdoctoral fellow supported
by the Consejería de Transformación Económica, Industria,
Conocimiento y Universidades de la Junta de Andalucía.

Sandra Catalán was supported by the grant RYC2021-033973-I, funded by MCIN/AEI/10.13039/501100011033 and the European Union "NextGenerationEU"/PRTR.

\end{acks}

\bibliographystyle{ACM-Reference-Format} 
\bibliography{biblio,hpc}


\begin{thebibliography}{22}


\ifx \showCODEN    \undefined \def \showCODEN     #1{\unskip}     \fi
\ifx \showDOI      \undefined \def \showDOI       #1{#1}\fi
\ifx \showISBNx    \undefined \def \showISBNx     #1{\unskip}     \fi
\ifx \showISBNxiii \undefined \def \showISBNxiii  #1{\unskip}     \fi
\ifx \showISSN     \undefined \def \showISSN      #1{\unskip}     \fi
\ifx \showLCCN     \undefined \def \showLCCN      #1{\unskip}     \fi
\ifx \shownote     \undefined \def \shownote      #1{#1}          \fi
\ifx \showarticletitle \undefined \def \showarticletitle #1{#1}   \fi
\ifx \showURL      \undefined \def \showURL       {\relax}        \fi
\providecommand\bibfield[2]{#2}
\providecommand\bibinfo[2]{#2}
\providecommand\natexlab[1]{#1}
\providecommand\showeprint[2][]{arXiv:#2}

\bibitem[Alaejos et~al\mbox{.}(2023)]%
        {Ala22}
\bibfield{author}{\bibinfo{person}{Guillermo Alaejos},
  \bibinfo{person}{Adri\'an Castell\'o}, \bibinfo{person}{H\'ector Mart\'inez},
  \bibinfo{person}{Pedro Alonso}, \bibinfo{person}{Francisco Igual}, {and}
  \bibinfo{person}{Enrique~S. Quintana-Ort\'i}.}
  \bibinfo{year}{2023}\natexlab{}.
\newblock \showarticletitle{Micro-kernels for portable and efficient matrix
  multiplication in deep learning}.
\newblock \bibinfo{journal}{\emph{The Journal of Supercomputing}}
  (\bibinfo{year}{2023}), \bibinfo{pages}{1--24}.
\newblock
\urldef\tempurl%
\url{https://doi.org/10.1007/s11227-022-05003-3}
\showDOI{\tempurl}


\bibitem[Anderson et~al\mbox{.}(1999)]%
        {lapack}
\bibfield{author}{\bibinfo{person}{Edward Anderson}, \bibinfo{person}{Zhaojun
  Bai}, \bibinfo{person}{L.~Susan Blackford}, \bibinfo{person}{James Demmel},
  \bibinfo{person}{Jack~J. Dongarra}, \bibinfo{person}{Jeremy~Du Croz},
  \bibinfo{person}{Sven Hammarling}, \bibinfo{person}{Anne Greenbaum},
  \bibinfo{person}{Alan McKenney}, {and} \bibinfo{person}{Danny~C. Sorensen}.}
  \bibinfo{year}{1999}\natexlab{}.
\newblock \bibinfo{booktitle}{\emph{{LAPACK} Users' guide}
  (\bibinfo{edition}{3rd} ed.)}.
\newblock \bibinfo{publisher}{SIAM}.
\newblock


\bibitem[Catal{\'a}n et~al\mbox{.}(2016)]%
        {Catalan2016}
\bibfield{author}{\bibinfo{person}{Sandra Catal{\'a}n},
  \bibinfo{person}{Francisco~D. Igual}, \bibinfo{person}{Rafael Mayo},
  \bibinfo{person}{Rafael Rodr{\'i}guez-S{\'a}nchez}, {and}
  \bibinfo{person}{Enrique~S. Quintana-Ort{\'i}}.}
  \bibinfo{year}{2016}\natexlab{}.
\newblock \showarticletitle{Architecture-aware configuration and scheduling of
  matrix multiplication on asymmetric multicore processors}.
\newblock \bibinfo{journal}{\emph{Cluster Computing}} \bibinfo{volume}{19},
  \bibinfo{number}{3} (\bibinfo{year}{2016}), \bibinfo{pages}{1037--1051}.
\newblock
\showISSN{1573-7543}
\urldef\tempurl%
\url{https://doi.org/10.1007/s10586-016-0611-8}
\showDOI{\tempurl}


\bibitem[Demmel(1997)]%
        {doi:10.1137/1.9781611971446}
\bibfield{author}{\bibinfo{person}{James~W. Demmel}.}
  \bibinfo{year}{1997}\natexlab{}.
\newblock \bibinfo{booktitle}{\emph{Applied Numerical Linear Algebra}}.
\newblock \bibinfo{publisher}{Society for Industrial and Applied Mathematics}.
\newblock


\bibitem[Dongarra et~al\mbox{.}(1990)]%
        {BLAS3}
\bibfield{author}{\bibinfo{person}{Jack~J. Dongarra}, \bibinfo{person}{Jeremy
  Du~Croz}, \bibinfo{person}{Sven Hammarling}, {and} \bibinfo{person}{Iain
  Duff}.} \bibinfo{year}{1990}\natexlab{}.
\newblock \showarticletitle{A Set of Level 3 Basic Linear Algebra Subprograms}.
\newblock \bibinfo{journal}{\emph{ACM Trans. Math. Softw.}}
  \bibinfo{volume}{16}, \bibinfo{number}{1} (\bibinfo{year}{1990}),
  \bibinfo{pages}{1--17}.
\newblock
\urldef\tempurl%
\url{https://doi.org/10.1145/77626.79170}
\showDOI{\tempurl}


\bibitem[Dongarra et~al\mbox{.}(1998)]%
        {DonDSV98}
\bibfield{author}{\bibinfo{person}{Jack~J. Dongarra}, \bibinfo{person}{Iain~S.
  Duff}, \bibinfo{person}{Danny~C. Sorensen}, {and} \bibinfo{person}{Henk~A.
  van~der Vorst}.} \bibinfo{year}{1998}\natexlab{}.
\newblock \bibinfo{booktitle}{\emph{{Numerical Linear Algebra for
  High-Performance Computers}}}.
\newblock \bibinfo{publisher}{Society for Industrial and Applied Mathematics}.
\newblock


\bibitem[Dowd and Severance(1998)]%
        {Dowd98}
\bibfield{author}{\bibinfo{person}{Kevin Dowd} {and}
  \bibinfo{person}{Charles~R. Severance}.} \bibinfo{year}{1998}\natexlab{}.
\newblock \bibinfo{booktitle}{\emph{High Performance Computing}
  (\bibinfo{edition}{2nd} ed.)}.
\newblock \bibinfo{publisher}{O'Reilly}.
\newblock


\bibitem[Golub and Ortega(1991)]%
        {GO91}
\bibfield{author}{\bibinfo{person}{Gene~H. Golub} {and}
  \bibinfo{person}{Jammes~M. Ortega}.} \bibinfo{year}{1991}\natexlab{}.
\newblock \bibinfo{booktitle}{\emph{Scientific Computing and Differential
  Equations. An Introduction to Numerical Methods}}.
\newblock \bibinfo{publisher}{Academic Press}.
\newblock


\bibitem[Golub and {Van~Loan}(1996)]%
        {GVL3}
\bibfield{author}{\bibinfo{person}{Gene~H. Golub} {and}
  \bibinfo{person}{Charles~F. {Van~Loan}}.} \bibinfo{year}{1996}\natexlab{}.
\newblock \bibinfo{booktitle}{\emph{Matrix Computations}
  (\bibinfo{edition}{3rd} ed.)}.
\newblock \bibinfo{publisher}{The Johns Hopkins University Press},
  \bibinfo{address}{Baltimore}.
\newblock


\bibitem[Goto and {v}an~de Geijn(2008)]%
        {Goto:2008:HIL3}
\bibfield{author}{\bibinfo{person}{Kazushige Goto} {and}
  \bibinfo{person}{Robert {v}an~de Geijn}.} \bibinfo{year}{2008}\natexlab{}.
\newblock \showarticletitle{High-performance implementation of the level-3
  {BLAS}}.
\newblock \bibinfo{journal}{\emph{ACM Trans. Math. Soft.}}
  \bibinfo{volume}{35}, \bibinfo{number}{1} (\bibinfo{year}{2008}),
  \bibinfo{pages}{1--14}.
\newblock
\urldef\tempurl%
\url{https://doi.org/10.1145/1377603.1377607}
\showDOI{\tempurl}


\bibitem[Goto and van~de Geijn(2008)]%
        {Goto:2008:AHP}
\bibfield{author}{\bibinfo{person}{Kazushige Goto} {and}
  \bibinfo{person}{Robert~A. van~de Geijn}.} \bibinfo{year}{2008}\natexlab{}.
\newblock \showarticletitle{Anatomy of a High-Performance Matrix
  Multiplication}.
\newblock \bibinfo{journal}{\emph{{ACM} Trans. Math. Softw.}}
  \bibinfo{volume}{34}, \bibinfo{number}{3} (\bibinfo{year}{2008}),
  \bibinfo{pages}{12:1--12:25}.
\newblock
\urldef\tempurl%
\url{https://doi.org/10.1145/1356052.1356053}
\showDOI{\tempurl}


\bibitem[Hennessy and Patterson(2017)]%
        {HenP17}
\bibfield{author}{\bibinfo{person}{John~L. Hennessy} {and}
  \bibinfo{person}{David~A. Patterson}.} \bibinfo{year}{2017}\natexlab{}.
\newblock \bibinfo{booktitle}{\emph{Computer Architecture: A Quantitative
  Approach} (\bibinfo{edition}{5} ed.)}.
\newblock \bibinfo{publisher}{Morgan Kaufmann Pub.}, \bibinfo{address}{San
  Francisco}.
\newblock


\bibitem[K\r{a}gstr\"{o}m et~al\mbox{.}(1998)]%
        {10.1145/292395.292412}
\bibfield{author}{\bibinfo{person}{Bo K\r{a}gstr\"{o}m}, \bibinfo{person}{Per
  Ling}, {and} \bibinfo{person}{Charles van Loan}.}
  \bibinfo{year}{1998}\natexlab{}.
\newblock \showarticletitle{{GEMM}-Based Level 3 {BLAS}: High-Performance Model
  Implementations and Performance Evaluation Benchmark}.
\newblock \bibinfo{journal}{\emph{ACM Trans. Math. Softw.}}
  \bibinfo{volume}{24}, \bibinfo{number}{3} (\bibinfo{year}{1998}),
  \bibinfo{pages}{268–302}.
\newblock
\showISSN{0098-3500}
\urldef\tempurl%
\url{https://doi.org/10.1145/292395.292412}
\showDOI{\tempurl}


\bibitem[Low et~al\mbox{.}(2016)]%
        {BLIS4}
\bibfield{author}{\bibinfo{person}{Tze~Meng Low}, \bibinfo{person}{Francisco~D.
  Igual}, \bibinfo{person}{Tyler~M. Smith}, {and} \bibinfo{person}{Enrique~S.
  Quintana-Ort\'{\i}}.} \bibinfo{year}{2016}\natexlab{}.
\newblock \showarticletitle{Analytical Modeling Is Enough for High-Performance
  {BLIS}}.
\newblock \bibinfo{journal}{\emph{ACM Trans. Math. Softw.}}
  \bibinfo{volume}{43}, \bibinfo{number}{2}, Article \bibinfo{articleno}{12}
  (\bibinfo{year}{2016}), \bibinfo{numpages}{18}~pages.
\newblock
\showISSN{0098-3500}
\urldef\tempurl%
\url{https://doi.org/10.1145/2925987}
\showDOI{\tempurl}


\bibitem[McKee and Wisniewski(2011)]%
        {McKee2011}
\bibfield{author}{\bibinfo{person}{Sally~A. McKee} {and}
  \bibinfo{person}{Robert~W. Wisniewski}.} \bibinfo{year}{2011}\natexlab{}.
\newblock \bibinfo{booktitle}{\emph{Memory Wall}}.
\newblock \bibinfo{publisher}{Springer US}, \bibinfo{address}{Boston, MA},
  \bibinfo{pages}{1110--1116}.
\newblock
\showISBNx{978-0-387-09766-4}
\urldef\tempurl%
\url{https://doi.org/10.1007/978-0-387-09766-4\_234}
\showDOI{\tempurl}


\bibitem[Neon(2023)]%
        {neonweb}
Neon \bibinfo{year}{2023}\natexlab{}.
\newblock \bibinfo{title}{Learn the architecture -- {N}eon programmers' guide}.
\newblock
  \bibinfo{howpublished}{\url{https://developer.arm.com/documentation/den0018/a/NEON-Intrinsics}}.
\newblock
\newblock
\shownote{Last access: April 2023}.


\bibitem[{O}pen{BLAS}(2012)]%
        {OpenBLAS}
{O}pen{BLAS} \bibinfo{year}{2012}\natexlab{}.
\newblock \bibinfo{title}{{O}pen{BLAS}}.
\newblock \bibinfo{howpublished}{\url{http://xianyi.github.com/OpenBLAS/}}.
\newblock


\bibitem[Smith et~al\mbox{.}(2014)]%
        {BLIS3}
\bibfield{author}{\bibinfo{person}{Tyler~M. Smith}, \bibinfo{person}{Robert
  {van de Geijn}}, \bibinfo{person}{Mikhail Smelyanskiy},
  \bibinfo{person}{Jeff~R. Hammond}, {and} \bibinfo{person}{Field~G. {Van
  Zee}}.} \bibinfo{year}{2014}\natexlab{}.
\newblock \showarticletitle{Anatomy of High-Performance Many-Threaded Matrix
  Multiplication}. In \bibinfo{booktitle}{\emph{IPDPS '14: Proceedings of the
  International Parallel and Distributed Processing Symposium}}.
  \bibinfo{pages}{1049--1059}.
\newblock
\urldef\tempurl%
\url{https://doi.org/10.1109/IPDPS.2014.110}
\showDOI{\tempurl}


\bibitem[{Van~Zee} et~al\mbox{.}(2016)]%
        {BLIS2}
\bibfield{author}{\bibinfo{person}{Field~G. {Van~Zee}},
  \bibinfo{person}{Tyler~M. Smith}, \bibinfo{person}{Bryan Marker},
  \bibinfo{person}{Tze~Meng Low}, \bibinfo{person}{Robert~A. {Van~de~Geijn}},
  \bibinfo{person}{Francisco~D. Igual}, \bibinfo{person}{Mikhail Smelyanskiy},
  \bibinfo{person}{Xianyi Zhang}, \bibinfo{person}{Michael Kistler},
  \bibinfo{person}{Vernon Austel}, \bibinfo{person}{John~A. Gunnels}, {and}
  \bibinfo{person}{Lee Killough}.} \bibinfo{year}{2016}\natexlab{}.
\newblock \showarticletitle{The {BLIS} Framework: Experiments in Portability}.
\newblock \bibinfo{journal}{\emph{ACM Trans. Math. Softw.}}
  \bibinfo{volume}{42}, \bibinfo{number}{2}, Article \bibinfo{articleno}{12}
  (\bibinfo{date}{jun} \bibinfo{year}{2016}), \bibinfo{numpages}{19}~pages.
\newblock
\showISSN{0098-3500}
\urldef\tempurl%
\url{https://doi.org/10.1145/2755561}
\showDOI{\tempurl}


\bibitem[{Van~Zee} and {van~de~Geijn}(2015)]%
        {BLIS1}
\bibfield{author}{\bibinfo{person}{Field~G. {Van~Zee}} {and}
  \bibinfo{person}{Robert~A. {van~de~Geijn}}.} \bibinfo{year}{2015}\natexlab{}.
\newblock \showarticletitle{{BLIS}: A Framework for Rapidly Instantiating
  {BLAS} Functionality}.
\newblock \bibinfo{journal}{\emph{ACM Trans. Math. Softw.}}
  \bibinfo{volume}{41}, \bibinfo{number}{3} (\bibinfo{year}{2015}),
  \bibinfo{pages}{14:1--14:33}.
\newblock


\bibitem[Wulf and McKee(1995)]%
        {10.1145/216585.216588}
\bibfield{author}{\bibinfo{person}{Wm.~A. Wulf} {and} \bibinfo{person}{Sally~A.
  McKee}.} \bibinfo{year}{1995}\natexlab{}.
\newblock \showarticletitle{Hitting the Memory Wall: Implications of the
  Obvious}.
\newblock \bibinfo{journal}{\emph{SIGARCH Comput. Archit. News}}
  \bibinfo{volume}{23}, \bibinfo{number}{1} (\bibinfo{date}{mar}
  \bibinfo{year}{1995}), \bibinfo{pages}{20–24}.
\newblock
\showISSN{0163-5964}
\urldef\tempurl%
\url{https://doi.org/10.1145/216585.216588}
\showDOI{\tempurl}


\bibitem[Xianyi et~al\mbox{.}(2012)]%
        {OpenBLAS:ICPADS}
\bibfield{author}{\bibinfo{person}{Zhang Xianyi}, \bibinfo{person}{Wang Qian},
  {and} \bibinfo{person}{Zhang Yunquan}.} \bibinfo{year}{2012}\natexlab{}.
\newblock \showarticletitle{Model-driven Level 3 {BLAS} Performance
  Optimization on {L}oongson {3A} Processor}. In \bibinfo{booktitle}{\emph{2012
  IEEE 18th International Conference on Parallel and Distributed Systems
  ({ICPADS})}}.
\newblock


\end{thebibliography}

\end{document}